\documentclass[conference]{IEEEtran}
\IEEEoverridecommandlockouts
\usepackage{cite}
\usepackage[utf8]{inputenc}
\usepackage{graphicx}
\usepackage{amsmath}
\usepackage{amssymb}
\usepackage[version=4]{mhchem}
\usepackage{siunitx}
\usepackage{longtable,tabularx}
\usepackage{subcaption}
\usepackage{epstopdf}
\usepackage{tabularx}
\usepackage{booktabs} 
\usepackage{multirow}
\usepackage{arydshln} 
\usepackage{url}
\usepackage{float}
\usepackage{acronym}
\usepackage{fancyhdr}
\usepackage{threeparttable}
\usepackage{comment}
\usepackage{hyperref} 
\usepackage{textcomp} 
\usepackage{makecell}

\usepackage{hyperref} 
\usepackage{textcomp} 
\def\BibTeX{{\rm B\kern-.05em{\sc i\kern-.025em b}\kern-.08em
    T\kern-.1667em\lower.7ex\hbox{E}\kern-.125emX}}

\renewcommand{\footnoterule}{%
  \kern -3pt
  \hrule width \columnwidth height 1pt
  \kern 2pt
}

\begin{document}

\title{Quantifying Uncertainty in Space Debris Capture with Active Tether-Net Systems Caused by Noisy Observations}

\makeatletter
\newcommand{\linebreakand}{%
  \end{@IEEEauthorhalign}
  \hfill\mbox{}\par
  \mbox{}\hfill\begin{@IEEEauthorhalign}
}
\makeatother

\author{\IEEEauthorblockN{Feng Liu$^*$ \thanks{$^*$ Ph.D. Candidate, Department of Mechanical and Aerospace Engineering, AIAA Student Member}}
\IEEEauthorblockA{\textit{University at Buffalo}\\
Buffalo, New York, 14260 \\
fliu23@buffalo.edu}
\and
\IEEEauthorblockN{Achira Boonrath$^*$}
\IEEEauthorblockA{\textit{University at Buffalo}\\
Buffalo, New York, 14260 \\
achirabo@buffalo.edu}
\and
\IEEEauthorblockN{Eleonora M. Botta$^\dagger$ \thanks{$^\dagger$ Associate Professor, Department of Mechanical and Aerospace Engineering, AIAA Member}}
\IEEEauthorblockA{\textit{University at Buffalo}\\
Buffalo, New York, 14260 \\
ebotta@buffalo.edu}
\and
\IEEEauthorblockN{Souma Chowdhury$^\ddag$ \thanks{$^\ddag$ Professor, Department of Mechanical and Aerospace Engineering, AIAA Senior Member, Corr. author} 
\thanks{This work is accepted to be presented in the 2025 AIAA Aviation Forum.} 
}
\IEEEauthorblockA{\textit{University at Buffalo}\\
Buffalo, New York, 14260 \\
soumacho@buffalo.edu}
}

\maketitle

\newacro{LEO}{Low Earth Orbit}
\newacro{ADR}{Active Debris Removal}
\newacro{RL}{Reinforcement Learning}
\newacro{RNN}{Recurrent Neural Network}
\newacro{MU}{Maneuverable Unit}
\newacro{LSTM}{long-short term memory}
\newacro{CNN}{Convolutional Neural Network}
\newacro{QoI}{Quantity of Interest}
\newacro{LHS}{Latin Hypercube Sampling}
\newacro{KL}{Kullback-Leibler}
\newacro{RWM}{Random Walk Metropolis}
\newacro{CoM}{center of mass}
\newacro{MLE}{Maximum Likelihood Estimator}
\newacro{UQ}{uncertainty quantification}
\newacro{CQI}{Capture Quality Index}
\newacro{CH}{Convex Hull}
\newacro{MLP}{Multilayer Perceptron}

\thispagestyle{plain}
\pagestyle{plain}


\begin{abstract}
As Low Earth Orbit has grown more crowded with space debris, the need for reliable and efficient debris removal solutions becomes more urgent. An active tether-net system with maneuverable units is one of the promising solutions to this problem, whose success is dependent on the robustness of the net maneuver and closing decisions. These in turn are impacted by the uncertainties attributed to i) noisy observation of the target debris state (e.g., sensing errors), and ii) imperfect simulations of the complex net dynamics and net/debris interaction behavior, over which the decision system is trained. This paper focuses on the first of these two uncertainty sources, and presents a pipeline to propagate and quantify the resulting uncertainty in the debris capture performance expressed in terms of Capture Quality Index (CQI). This quantification is uniquely performed for both an active tether-net using a fixed baseline control and one using a trained neuro-control policy to guide the net maneuver during the deployment phase. Two different uncertainty quantification (UQ) techniques, namely Sobol's variance-based sensitivity analysis and perturbation-based method are exploited. A high-fidelity simulator and a lower-fidelity surrogate-based environment are used to demonstrate trade-offs between prediction accuracy versus ease of resolving uncertainties. 
\end{abstract}




\section{Introduction}
The increasing amount of space debris in \ac{LEO} poses a significant threat to the operational safety of current and future space missions \cite{colvin2023cost}, creating an urgent need for mitigation solutions. \ac{ADR} demonstrates the potential to alleviate the space debris problem \cite{lg_remove}, and among the proposed \ac{ADR} methods, tether-net systems prove to be effective for the capture of uncooperative debris with their high flexibility and relatively long capture range \cite{boonrath2025robustness, flex}. However, as net-based \ac{ADR} systems have relatively low technological maturity, additional research is needed to obtain a better understanding regarding the behavior of the system prior to employment in real debris removal missions.

Previous studies have researched the dynamics of net deployment (i.e., net launch and flight to the target) and debris capture processes using simulation tools employing physics-based models; these works include the research conducted by Medina et al. \cite{medina2017validation}, Benvenuto et al. \cite{benvenuto2015dynamics}, Shan et al. \cite{shan2019contact}, Si et al. \cite{Si2019CaptureSelfCollision2019}, Botta et al. \cite{botta2016simulation, botta2017energy, botta2019simulation}, Endo et al. \cite{endo2020study, endo2022new, kojima2023design}, Hou et al. \cite{hou2021dynamic}, and Huang et al. \cite{huang2022nonlinear, huang2023contact}. Researchers have shown that the robustness and reliability of the tether-net systems can be enhanced by the use of active robotic systems \cite{ approach,huang2016impact, ZhaoYakun2020ISSM, boonrath2024learning, zhu2024multi}; for such systems, multiple \acp{MU} are installed on the corners or the edges of the net, enhancing the operation distance and flexibility. With the robotic \acp{MU}, the trajectories of the systems can be adjusted during deployment, allowing for the tracking of complex paths and permitting the system to achieve debris capture in challenging scenarios.

Space debris capture missions require rapid and reliable responses to dynamic conditions, such as uncertain debris positions, angular velocity, and limited maneuvering windows, with multiple studies conducted under some level of uncertainty \cite{huang2016impact,jsr2023Zeng,boonrath2025robustness}. Learning-based methods, such as \ac{RL}, have emerged as promising solutions to enable control in such scenarios due to their faster execution time and robustness to varying mission environments compared to traditional control methods (e.g., state-feedback-based control algorithms). \ac{RL} can formulate generalized and optimized control actions in various scenarios without the direct need for a sufficiently accurate system model. Previous research \cite{zeng,liu2023learning} has, respectively, explored utilizing \ac{RL} to determine the net's closing condition and thrust controls of the \acp{MU}. However, \ac{RL} typically requires a large number of training samples for the development of a good policy model, which can still be computationally expensive and time-consuming to obtain, especially if complex high-fidelity simulations are required for the training. Thus, it is crucial to identify which variables and parameters in the space environment most significantly influence the \ac{QoI} of the capture mission, such as a capture success index and fuel consumption. Performing a sensitivity analysis allows for a better understanding of these critical factors, enabling more efficient sampling strategies and more focused, efficient policy learning for future studies. 
Once the \ac{RL} policy — represented as a neural network that determines the control of the \acp{MU} — is trained, it also becomes valuable to investigate how the sensitivity analysis can be applied to help further improve the \ac{RL} policy. Specific scenarios or regions of the input space can be identified where the policy performs poorly or fails to generalize, revealing potential weaknesses that may not be apparent during standard training evaluation. Such insights can guide targeted retraining efforts or policy refinement, ultimately improving the robustness and reliability of the control system under a wide range of environmental conditions.
 
Nonetheless, prior to conducting a sensitivity analysis, the high computational expense of high-fidelity simulators employing physics-based models (as used in the previously cited works) remains a significant challenge. The high computational costs of physics-based models can be attributed to the collision detection performed between the net, which often consists of $\mathcal{O}(10^2)$ small bodies interlinked with each other, and the debris at every simulation time step. 
This limitation restricts not only the number of simulations that can be performed within a reasonable time, but also the number of input variables that can be included in the analysis. As a result, it becomes challenging to conduct a thorough sensitivity evaluation across a large design space, potentially overlooking influential variables that impact mission success. Addressing this bottleneck is essential for enabling more comprehensive and effective learning-based optimization strategies. Therefore, employing neural-network-based, data-driven surrogate models to take over a portion or all of the simulation process presents an efficient alternative. Boonrath et al. proposed using a high-fidelity model to simulate a net-based debris capture mission only until the moment before net closing, and then use an \ac{RNN} network for predicting the capture results, saving approx. ~68\% computing time and demonstrating the successful application of a surrogate model for tether-net system simulations \cite{boonrath2024learning}. Thus, in this paper, we propose to simulate the full debris capture mission using multiple surrogate models for the \ac{UQ} analysis, with an analysis using the high-fidelity simulator also included as a comparison.

As a result of the inherent behavioral complexity of the tether-net capture system, regardless of whether a high-fidelity physics-based or a data-driven surrogate model is employed to simulate its dynamics, an analytical approach to \ac{UQ} and propagation becomes extremely challenging. Moreover, deriving closed-form sensitivity metrics for high-dimensional, nonlinear systems is often difficult. Because of the challenges, sampling-based estimators such as Sobol's variance-based sensitivity analysis using Saltelli's sampling scheme \cite{Saltelli_est} offer a more practical and scalable alternative. In comparison, perturbation-based methods require significantly fewer data to be sampled \cite{perturb, perturb_gw} whilst still being capable of yielding useful insights into local sensitivity characteristics. 
Therefore, these methods are valuable and can effectively estimate the contribution of individual variables and their interactions to output variability without relying on explicit analytical formulations, making them well-suited for use with surrogate models or high-fidelity simulations as a black-box of complex systems.

Given that so far there is no uncertainty analysis conducted for a simulated tether-net system where the high-fidelity model is fully replaced by a series of surrogate models - particularly in scenarios where a pre-trained \ac{RL} policy is embedded within the simulation framework -- the contributions of this work are summarized as follows: \textbf{1) Sensitivity analysis and uncertainty propagation framework:} We proposed an uncertainty analysis framework developed for the tether-net space debris capture system that works for multiple simulation fidelities. \textbf{2) Investigation on the influence of target position on the capture performance: } The influence of the initial target position—specifically along the X, Y, and Z axes—on \ac{CQI} \cite{CQI_Original, barnes} (which will be discussed in later Section) and fuel consumption is systematically investigated. \textbf{3) Analysis with various fidelity for both Fixed- and Active-Control Nets:} 
Two thrust control formulations are considered for the tether-net system in this work. First, for the \textit{Fixed-Control Net}, the thrust assigned for each \ac{MU} is unchanged across all considered scenarios. In contrast, for the \textit{Active-Control Net}, there is an embedded \ac{RL} policy that is used to assign the thrust for each \ac{MU}, depending on the states of the target in an \ac{ADR} scenario. Using these two control formulations, we conducted the analysis across three different simulation environments: a high-fidelity physics-based simulator with the Fixed-Control Net, a high-fidelity physics-based simulator with the Active-Control Net, and a surrogate-models-based simulator employing nets with both fixed and active control methods. \textbf{4) Analysis with perception uncertainty in the Active-Control Nets:} The input to the RL policy is added a noise, and we investigated the sensitivity of the Active-Control Net to the noisy perception in a scenario.

The paper is organized as follows: Section \ref{sec:sim_mod} introduces the modeling of the tether-net system and the \ac{QoI}, as well as the high-fidelity model, the \ac{RL}, and the surrogate model's inputs and outputs. Section \ref{sec:UQ_formula} introduces the \ac{UQ} methods used in the paper and the overall framework. Section \ref{sec:case_study} explains the case studies developed for the \ac{UQ}. Section \ref{sec:results} demonstrates and discusses the results. Section \ref{sec:conclude} presents concluding remarks. An Appendix is included to show the supplement results.

The robotic tether-net system used in this work is based on the design used by the previous work of Liu et al. \cite{liu2023learning}. As Fig.~\ref{fig:topen_tclose} shows, the tether-net system consists of a chaser satellite, a net that is connected to the chaser through a main tether, a winch-based closing mechanism, and four \acp{MU} equipped with thrusters. The inertial reference frame $\mathcal{O}=\{O,\mathbf{\hat{i}},\mathbf{\hat{j}},\mathbf{\hat{k}}\}$ and the target body-fixed frame $\mathcal{D}=\{D,\mathbf{\hat{d}}_x,\mathbf{\hat{d}}_y, \mathbf{\hat{d}}_z\}$, whose origins $O$ and $D$ are fixed to the initial location of the net's corner knot adjacent to \ac{MU}$_1$ and to the target's \ac{CoM}, respectively. The target to be captured has a cylindrical shape with the dimensions of 11 m length and 3.9 m diameter; it is approximately modeled after the Zenit-2 upper stage, which was the target of interest in the work of Botta et al. \cite{botta2019simulation}. In this work, the \ac{CoM} of the target is assumed to be the same as its geometric center. The target is assumed to be initially spinning, but not translating, in relation to the chaser spacecraft; its position, initial orientation, and angular velocity are assumed to vary between each of the capture scenarios considered. 

The net is initially stored on the chaser satellite and is ejected from the satellite after the chaser is at the launch location. After the launch, the net is left to open freely for 15 s; then, the thrusters on the \acp{MU} are turned on, maneuvering the net toward the target. The open-loop control of the tether-net system includes nine control variables: four thrust angles $\psi_{T,i, i=1,2,3,4}$ on $\mathbf{\hat{i}}$$\mathbf{\hat{j}}$ plane and four thrust angles $\theta_{T,i, i=1,2,3,4}$. on $\mathbf{\hat{j}}$$\mathbf{\hat{k}}$ plane, and thrust magnitude $F_T$ that is assumed to be identical for all four \acp{MU}. The thrusters on the \acp{MU} are assumed to be sufficiently precise and capable of providing the assigned directions and magnitudes of the commanded thrusts. Specifically, the thrusts are defined in the following Eq. \eqref{eq:thrusts1}. 
\begin{equation}
\begin{aligned}
\mathbf{F}_{T,i} ={}& \pm F_{T}\sin(\theta_{T,i}) \cos(\psi_{T,i})\mathbf{\hat{i}}
 \pm F_{T}\sin(\theta_{T,i})\sin(\psi_{T,i})\mathbf{\hat{j}}\\
& \quad\quad\quad - F_{T}\cos(\theta_{T,i})\mathbf{\hat{k}}
\end{aligned}
\label{eq:thrusts1}
\end{equation}
The definitions of the thrust angles relative to the $\mathcal{O}$ frame axes are illustrated in Fig. \ref{fig:control_plot}, noting that the thrust angles are defined such that the $\psi_{T,i}$ and $\theta_{T,i}$ positive angles and 0$^\circ$ directions depend on the specific $i$-th \ac{MU}. The information regarding the target states and \ac{MU} control is provided in Table \ref{tab:optbounds}. 

The time when the thrusters are turned on is referred to as $t_{\text{on}} = 15$ s. There are two more important instances after $t_{\text{on}}$, up to the time when the net closes. The first instance is defined as $t_{\text{open}}$, which represents the time when the \ac{CoM} of the four \acp{MU} passes the plane placed +5.5$\hat{\mathbf{k}}$ m away from the \ac{CoM} of the target and parallel to $\hat{\mathbf{i}}$$\hat{\mathbf{j}}$ (see the green plane in Fig. \ref{fig:topen_tclose}); it is an approximation of an instance soon \textbf{before} the net contacts the target, and the states of the \acp{MU} at this instance are critical to determine if the capture will be successful. The second important instance is $t_{\text{close}}$, which is the time that the \textbf{all} \acp{MU} passes the plane placed -5.5$\hat{\mathbf{k}}$ m away from the \ac{CoM} of the target (see the red plane in Fig. \ref{fig:topen_tclose}); it is also the time when all the thrusters are turned off and the closing mechanism is activated, so the thrusters of the \acp{MU} are turned on for a total of $t_{\text{close}}-t_{\text{on}}$ s. Also, $t_{\text{close}}$ is set to be at maximum 10 s after $t_{\text{on}}$, in case the \acp{MU} fail to satisfy the closing condition mentioned earlier. For this work, the time period between $t_{\text{on}}$ and $t_{\text{close}}$ is referred to as \textbf{deployment phase}, and the time period from $t_{\text{close}}$ till the end of the simulation is called \textbf{capture phase}. After 10 s following the activation of the closing mechanism, the final capture performance of the system is evaluated.

There are two values that should be optimized for in a space debris capture mission employing a robotic tether-net: the total fuel consumption of all the \acp{MU}, $G$, and \ac{CQI} value at the end of the mission, $J$. It is assumed that the fuel consumption rate is proportional to the thrust magnitude ${F}_{T}$, identical for all 4 \acp{MU}. The fuel consumption rate is defined as $0.0014{F}_{T}$ kg/s \cite{thrusters}, and the total fuel consumption is defined in Eq. \eqref{eq:fuel}.
\begin{equation}
 G = 4(0.0014)\left(t_{\text{close}} - t_{\text{on}}\right){F}_{T}
\label{eq:fuel}
\end{equation} 
The other value of importance is the \ac{CQI}: the \ac{CQI} reflects the similarity between the \ac{CH} shape of the net and the target, and the distance between their \acp{CoM}. The \ac{CQI} is defined as:
\begin{equation}
 J = 0.1\frac{|V_n-V_D|}{V_D}+0.1\frac{|S_n-S_D|}{S_D} +0.8\frac{|q_n|}{L_c}
 \label{eq:cqiSafe}
\end{equation}
The variables in the expression include the \ac{CH} volume of the net $V_n$, the target's volume $V_D$, the \ac{CH} surface area of the net $S_n$, the surface area of the target $S_D$, the distance between the \ac{CoM} of the target and the \ac{CoM} of the net $q_n$, and the minimum distance from the target's \ac{CoM} to its surface denoted as $L_c$. In previous work \cite{liu2023learning, boonrath2024learning}, a capture with a \ac{CQI} of less than $2.5$ and a number of locked pairs greater than 10 at the final simulation time is defined as a successful capture. However, in this paper, we will focus only on the values of fuel consumption and \ac{CQI} for the \ac{UQ}. 


\begin{figure*}[htp!]
 \centering
 \includegraphics[width=0.7\textwidth]{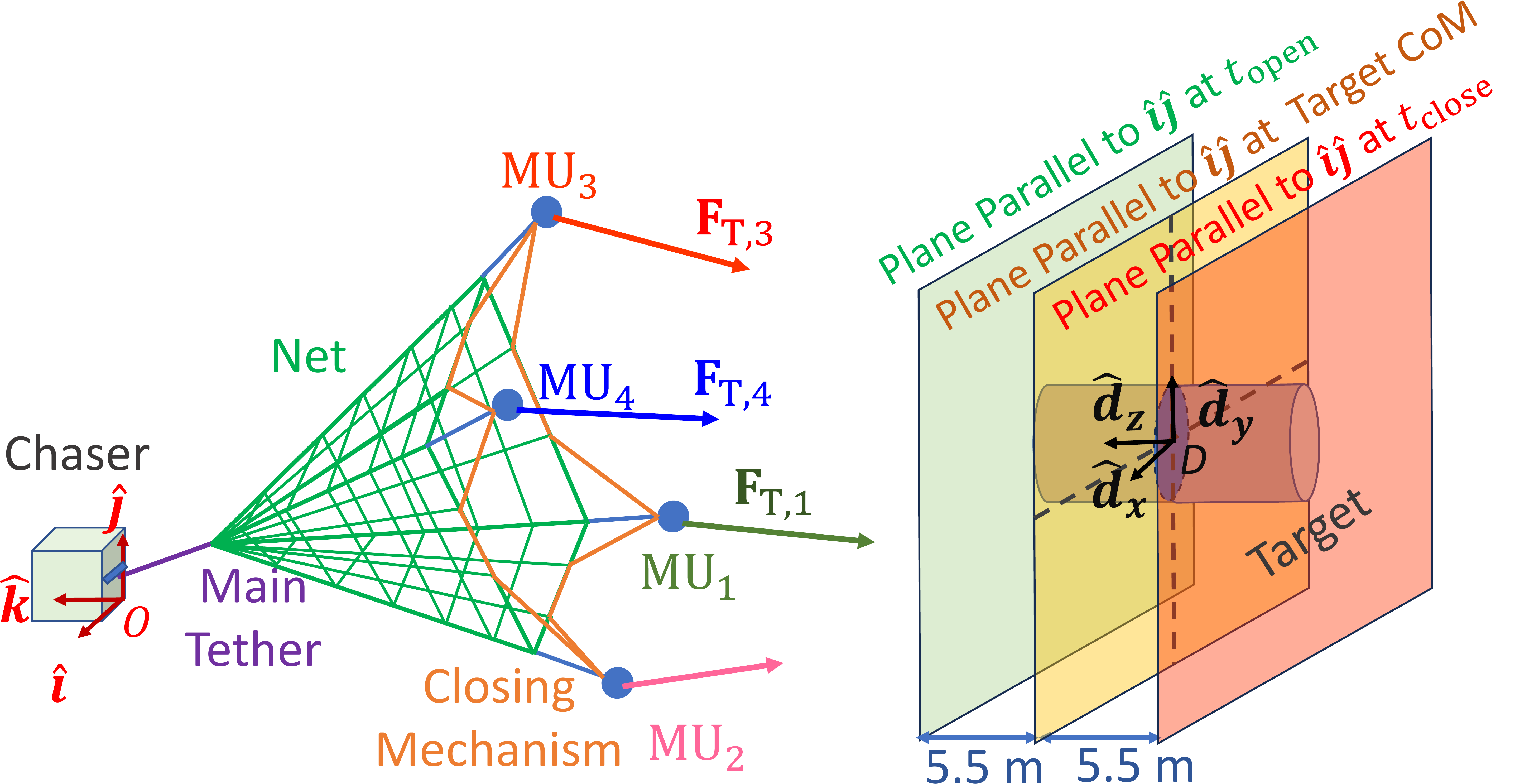}
 \caption{Tether-Net System for Debris Capture}
 \label{fig:topen_tclose}
\end{figure*}

\begin{figure}[htp!]
 \centering
 \includegraphics[width=1.0\linewidth]{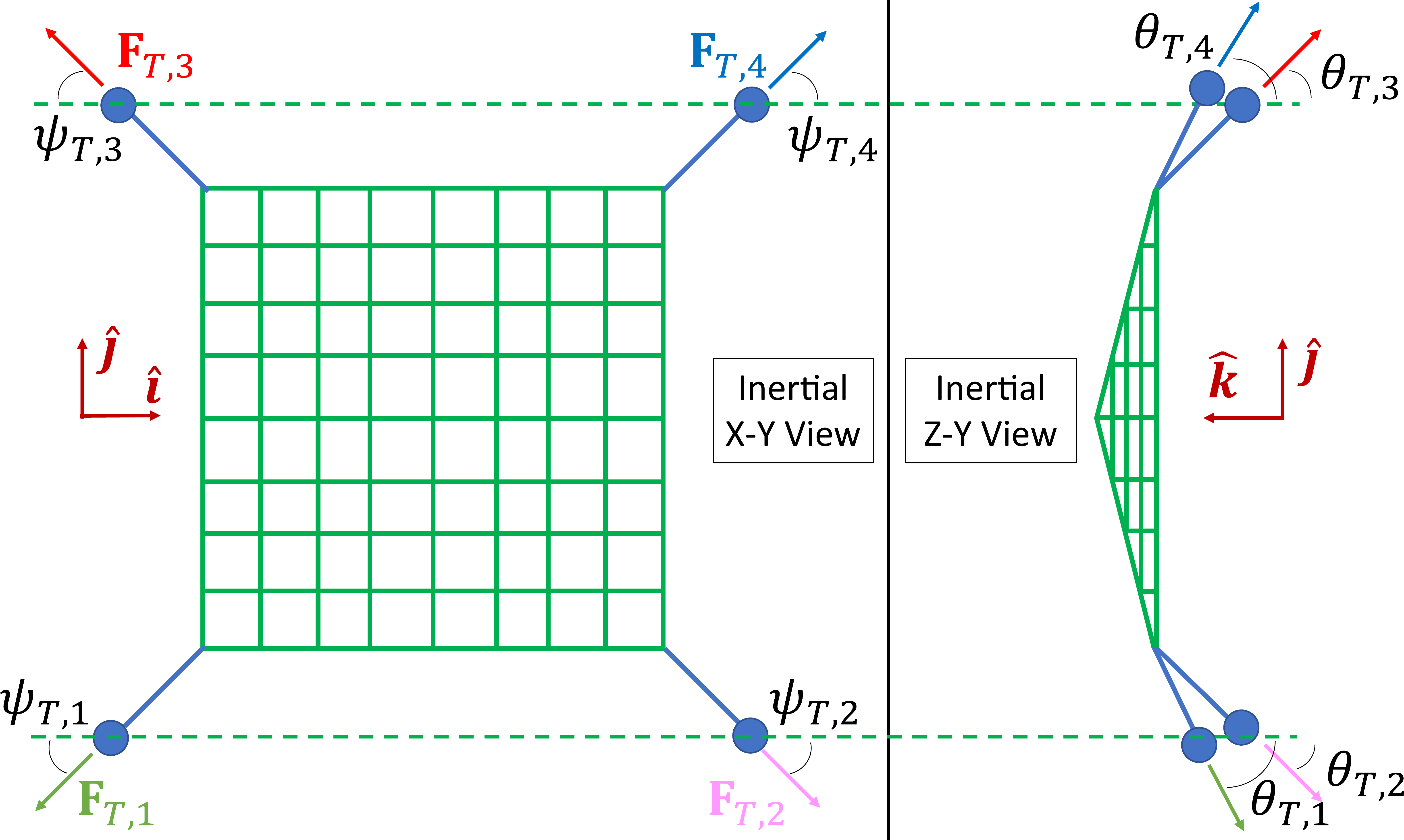}
 \caption{Thrust angles for the \ac{MU}}
 \label{fig:control_plot}
\end{figure}


\begin{table*}[htp!]
 \centering
 \begin{threeparttable}
 \caption{Simulation Input Variables Definitions and Bounds}
 \label{tab:optbounds}
 \begin{tabular}{lllll}
 \toprule
 & Variables & Description & Bounds & Unit \\
 \midrule
 \multirow{9}{*}{Target State} 
& $X_D$ & target position coordinate along the $\boldsymbol{\hat{i}}$ direction & -9 to 9 & m\\
& $Y_D$ & target position coordinate along the $\boldsymbol{\hat{j}}$ direction & -9 to 9 & m\\
& $Z_D$ & target position coordinate along the $\boldsymbol{\hat{k}}$ direction & -60 to -40 & m\\
& $O_{D_{x0}}$ & x-component of the 321-Euler Angles set of $\mathcal{D}$ & -180.0 to 180.0 & deg \\
& $O_{D_{u0}}$ & y-component of the 321-Euler Angles set of $\mathcal{D}$ & -180.0 to 180.0 & deg \\
& $O_{D_{z0}}$ & z-component of the 321-Euler Angles of $\mathcal{D}$ & -180.0 to 180.0 & deg \\
& $\omega_{D_{x0}}$ & x-component of the angular velocity of $\mathcal{D}$ & 1.0 to 10.0 & deg/s \\
& $\omega_{D_{y0}}$ & y-component of the angular velocity of $\mathcal{D}$ & 1.0 to 10.0 & deg/s \\
& $\omega_{D_{z0}}$ & z-component of the angular velocity of $\mathcal{D}$ & 5.0 to 30.0 & deg/s \\
 \midrule
 \multirow{3}{*}{MU Control} & $\psi_{T,i, \: i=1,2,3,4}$ & Thrust angle on inertial $\boldsymbol{\hat{i}}$$\boldsymbol{\hat{j}}$ plane for the MU & -180 to 180 & deg\\
& $\theta_{T,i, \: i=1,2,3,4}$ & Thrust angle on inertial $\boldsymbol{\hat{j}}$$\boldsymbol{\hat{k}}$ plane for the MU & 35 to 55 & deg\\
 & $F_{T}$ & Thrust magnitude & 5 to 12 & N\\
 \bottomrule 
 \end{tabular}
 \end{threeparttable}
\end{table*}

\section{Models and RL Control Policy for the Tether-Net System}\label{sec:sim_mod}

For the system described in the previous section, we developed a high-fidelity physics-based simulator, serving as the ground truth for evaluating the performance of the tether-net system. To lower the computational burden while preserving essential dynamics, we also developed surrogate models for both the deployment and capture phases separately. These surrogate models enable rapid evaluations and are particularly useful for \ac{UQ}, especially when a large number of samples is necessary. We also integrated an \ac{RL} policy in the simulator to pick the \ac{MU} control parameters given the target states (see Table \ref{tab:optbounds} defines the bounds of the target states and \ac{MU} control parameters). The following text will discuss the high-fidelity simulator, the surrogate models, and the \ac{RL} policy. Note that as the focus of the paper is mainly on the \ac{UQ} framework, details regarding the training of the \ac{RL} and the surrogate models will not be included. The workstation used in this research has an AMD Ryzen 9 5950X- 16-Core processor, 64 GB RAM, and 12 GB NVIDIA GeForce RTX- 3060 GPU. The computational times mentioned in the rest of the paper are based on this hardware setup.

\subsection{High-Fidelity Simulator}
Driven by the goals of this research, the simulation tool for this work was created using Python, enabling efficient deployment on high-performance computing resources, such as Linux-based supercomputers. The simulator utilizes Google's \texttt{jax.numpy.array} data structure, which provides significant performance enhancements in numerical computation compared to the widely-used \texttt{numpy.array} \cite{jax2018github}. In combination with Google JAX, the Diffrax library supplies a variety of numerical integration methods to support the simulator's functionality \cite{kidger2022neural}. The selected numerical solver used to solve the equations of motion for the described system is \texttt{diffrax.Bosh3}, an implementation of the Bogacki-Shampine 3/2 adaptive timestep Runge-Kutta algorithm \cite{kidger2022neural}. The high-fidelity simulator is utilized to generate data for training and testing of the surrogate models. The computation time for a single simulation sample ranges from 6 to 15 minutes and can increase significantly if the net completely misses the target, resulting in a numerically unstable simulation. 
Within a simulation, the capture phase takes more than twice the computing time of the deployment phase to finish. Therefore, surrogate models are beneficial, saving a significant amount of computing time. The input vector to the simulator consists of the target states and the \ac{MU} control parameters in Table \ref{tab:optbounds}, and the output of the simulator includes the final \ac{CQI}, total fuel consumptions, and the dynamic record of the net nodes, target, and \acp{MU}.

\subsection{Surrogate Models for Deployment Phase and Capture Phase}
Like the high-fidelity simulator, the surrogate models are also Python-based, allowing for a consistent software implementation for the framework. Since it is challenging to directly predict capture results directly from the simulation initialization, 
we decoupled the process by using a deployment phase surrogate model and two capture phase models. A pre-trained \ac{LSTM} network serves as the deployment surrogate model. The details (number of neurons, layers, etc.) of the models are the content of another parallel paper, and will be posted to GitHub once the paper is accepted. The structure of the model is shown in Fig. \ref{fig:lstm_struct}. The target states and thrust parameters are the inputs that initialize the hidden states. These inputs are also duplicated into a sequence of vectors as the $X_t$ for each time step. At each time step, the input hidden states and the cell states are the outputs from the previous time step, and the output hidden state $h_t$ is a twelve-element vector representing the normalized relative displacement of $\boldsymbol{\hat{i}}$$\boldsymbol{\hat{j}}$$\boldsymbol{\hat{k}}$ coordinates of the four \acp{MU} from $t_{\text{on}}$ till $t_{\text{close}}$. An example of the current \ac{LSTM} model prediction result compared with the ground truth trajectories is shown in Fig. \ref{fig:lstm-traj}.
\begin{figure}[htp!]
 \centering
 \includegraphics[width=0.8\linewidth]{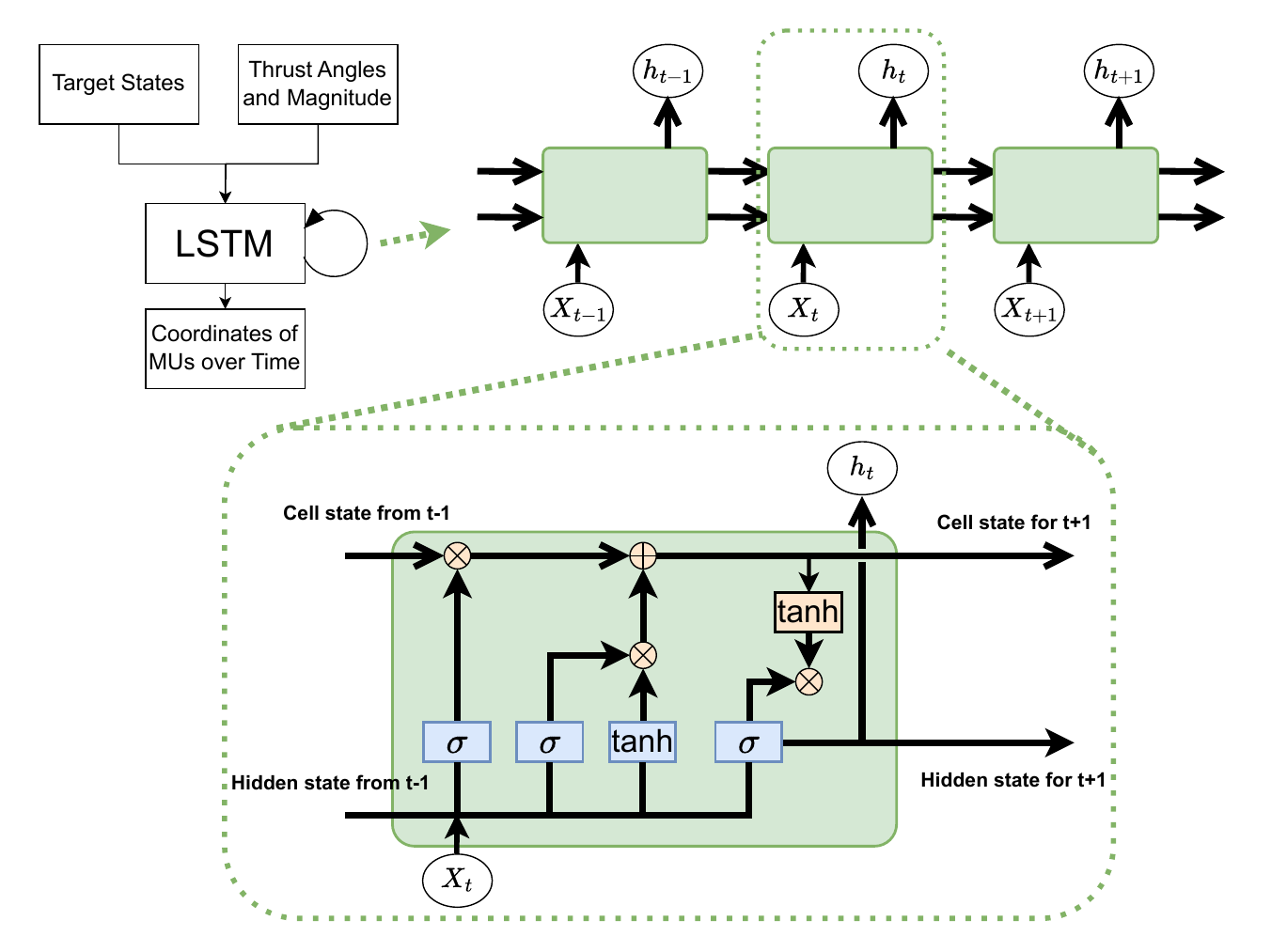}
 \caption{Deployment Phase Surrogate Model Structure}
 \label{fig:lstm_struct}
\end{figure}

\begin{figure}[htp!]
 \centering
 \includegraphics[width=0.8\linewidth]{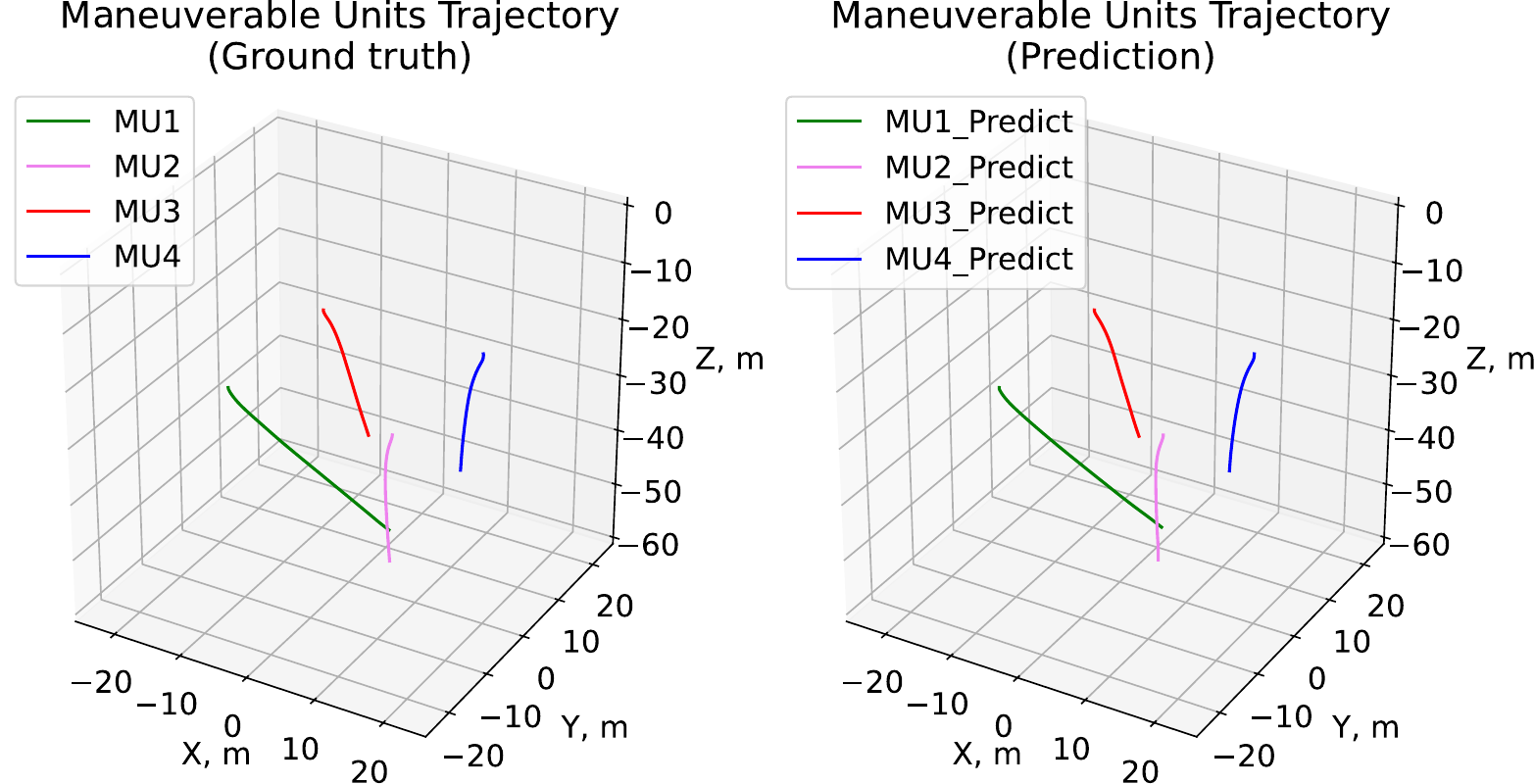}
 \caption{\ac{LSTM} Predicted Trajectories Comparison Example}
 \label{fig:lstm-traj}
\end{figure}
For the capture phase, two \acp{MLP} are trained for predicting the total fuel consumption $G$ and the final \ac{CQI} $J$. The two \acp{MLP} have the same structure as Table \ref{tab:mlp_structure} illustrated. The input to both \acp{MLP} is a vector consisting of target states and the position and velocity of each \ac{MU} at $t_{\text{open}}$. The \ac{MU} position and velocity are recalculated from the trajectory recreated from the output of the deployment phase surrogate model. Since the surrogate models in both phases will be used together, training with the \ac{LSTM}-predicted \acp{MU} states and ground-truth (from the high-fidelity simulator) target states and \acp{QoI} is more beneficial, because the model takes the error of \ac{LSTM} into consideration as well. The \ac{MLP} for the total fuel used the Mean Squared Error loss function for the training. For the \ac{CQI}, its value is transformed to log scale as log(1 + $J$), and the loss function used for training is the Smoothed L1 loss since the distribution of the \ac{CQI} is not balanced in the collected dataset. Each \ac{MLP} was trained with 6,864 samples and validated with 500 samples. The training and validation losses of the two \acp{MLP} are shown in Fig. \ref{fig:loss}. 

\begin{figure}[htp!]
 \centering
 \includegraphics[width=1.0\linewidth]{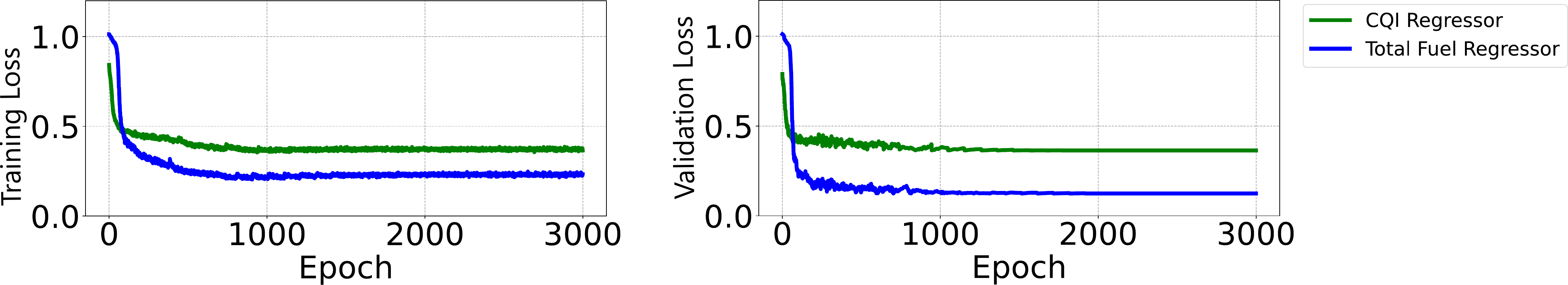}
 \caption{Capture Phase Loss}
 \label{fig:loss}
\end{figure}

\begin{table}[ht]
\centering
\footnotesize
\caption{Architecture of the Capture Phase MLP}
\label{tab:mlp_structure}
\begin{tabular}{|c|c|c|}
\hline
\textbf{Layer \#} & \textbf{Layer Type} & \textbf{Details} \\
\hline
1 & Linear & Input: \texttt{input\_size}, Output: 256 \\
2 & ReLU & -- \\
3 & Dropout & $p=0.5$ \\
\hline
4 & Linear & Input: 256, Output: 128 \\
5 & ReLU & -- \\
6 & Dropout & $p=0.5$ \\
\hline
7 & Linear & Input: 128, Output: 64 \\
8 & ReLU & -- \\
9 & Dropout & $p=0.5$ \\
\hline
10 & Linear & Input: 64, Output: 32 \\
11 & ReLU & -- \\
12 & Dropout & $p=0.5$ \\
\hline
13 & Linear & Input: 32, Output: 16 \\
14 & ReLU & -- \\
15 & Dropout & $p=0.5$ \\
\hline
16 & Linear & Input: 16, Output: 1 \\
\hline
\end{tabular}
\end{table}

\subsection{RL Policy for MU Control}
To explore the \ac{UQ} with an \ac{RL} policy network used as a thrust controller, we used a pre-trained \ac{RL} policy network from a parallel research. The policy was trained with the proximal policy optimization algorithm of Stable-Baselines3 implementation \cite{ppo}. The policy network is an \ac{MLP} that takes the target states as inputs and returns the thrust angles and magnitude as outputs. The \ac{RL} policy was trained with a reward function that was formulated to encourage \ac{MU} control that minimizes fuel consumption while achieving a successful capture. The trained policy has been validated with the high-fidelity simulator with 1,000 samples with random target states, achieving a successful capture rate of 85.2\% and a median fuel consumption of 0.172 kg. We will not expand the details of the \ac{RL} policy network, since for this work it is used as a black-box controller embedded in the simulation framework, either employed in the high-fidelity or surrogate-model-based simulations. The design details of experiments using this \ac{RL} will be illustrated in Section \ref{sec:case_study}, and the \ac{UQ} can show the influence of the \ac{RL} on the \acp{QoI} in Section \ref{sec:results}.

\section{Sensitivity Analysis and Uncertainty Propagation Framework}\label{sec:UQ_formula}
Sobol indices are a powerful tool for \ac{UQ}, as they help identify the relative influence of each input variable on the output response. By decomposing the output variance, Sobol indices provide valuable insight into which parameters contribute most to uncertainty, enabling more informed modeling and decision-making. To simplify the analysis in this study, we focus on varying only the target’s initial X, Y, and Z coordinates, referred to as $\mathbf{P} = [X_D, Y_D, Z_D]^T$, which are considered the primary sources of uncertainty. We also used a perturbation method to investigate the influence of varying target positions with a given standard deviation.

The sensitivity indices can be more efficiently acquired using the sampling-heavy Monte-Carlo estimation approach proposed by Saltelli \cite{SaltelliAndrea2002Mbuo} and Sobol \cite{SobolI} with a subset of the collected dataset. To estimate the first-order sensitivity indices and the total effect indices of the variables in Table \ref{tab:optbounds}, two matrices with $M \times p$ elements are created,

\begin{equation}
\begin{aligned}
\mathbf{A}=&\begin{bmatrix}\eta_1^{(1)}&\cdots&\eta_i^{(1)}&\cdots&\eta_p^{(1)}\\
\vdots&&&&\vdots\\
\eta_1^{(M)}&\cdots&\eta_i^{(M)}&\cdots&\eta_p^{(M)}
\end{bmatrix},\\
\mathbf{B}=&\begin{bmatrix}\hat\eta_1^{(1)}&\cdots&\hat\eta_i^{(1)}&\cdots&\hat\eta_p^{(1)}\\
\vdots&&&&\vdots\\
\hat\eta_1^{(M)}&\cdots&\hat\eta_i^{(M)}&\cdots&\hat\eta_p^{(M)}\end{bmatrix}
\end{aligned}
\end{equation}
where $M=500$ is the sample number of the subset of the original dataset, $p=3$ is the number of variables, $\eta_i$ is the input variables on the $i^{th}$ element of $\mathbf{P}$, $\eta_i^{(j)}$ and $\hat\eta_i^{(j)}$ are two samples generated from the uniform distribution of $\eta_i$, which are uniformly sampled from the original dataset.

Then, the matrix $\mathbf{C}_i$ is created:
\begin{equation}
\mathbf{C}_i=\begin{bmatrix}\hat{\eta}_1^{(1)}&\cdots&\eta_i^{(1)}&\cdots&\hat{\eta}_p^{(1)}\\
\vdots&&&\vdots\\
\hat{\eta}_1^{(M)}&\cdots&\eta_i^{(M)}&\cdots&\hat{\eta}_p^{(M)}\end{bmatrix},\quad i=1,\cdots,p
\end{equation}
in which each $\mathbf{C}_i$ is identical to $\mathbf{B}$ with the exception that the $i^{th}$ column is replaced by the $i^{th}$ column from $\mathbf{A}$. The collection of all the $\mathbf{C}_i$ is defined to be $\mathcal{C}$.

Each row of the above three matrices represents the inputs to the simulator, and the \ac{QoI} are the positions of \acp{MU} at \(t_\text{open}\) produced by the simulator, which is a vector of twelve elements, each representing an inertial position coordinate for the \acp{MU}. For matrices $\mathbf{A}$, $\mathbf{B}$ and $\mathcal{C}$, the \ac{QoI} is saved in the following variables, where $F$ represents the operations of executing function evaluation (the simulator or surrogate models) and fetching the \ac{QoI} from the output. To run the function, the target angular velocity vector $\mathbf{\Omega} = [\omega_{D_{x0}}, \omega_{D_{y0}}, \omega_{D_{z0}}]^T$, initial orientation vector $\mathbf{O} = [O_{D_{x0}}, O_{D_{y0}}, O_{D_{z0}}]^T$, and \ac{MU} control vector $\mathbf{X}$ are also needed. The values of $\mathbf{\Omega}$ and $\mathbf{O}$ are set as their mean values based on the bounds in Table \ref{tab:optbounds}, and the value of $\mathbf{X}$ is set as the mean value unless the \ac{RL} policy is applied to determine the control. For simplicity, we define a vector $\mathbf{D} = [\mathbf{\Omega}^T, \mathbf{O}^T, \mathbf{X}^T ]^T$ to be the collection of these variables.
 
\begin{equation}
\begin{aligned}
\mathbf{y}_A=F(\mathbf{A, D}),\quad\mathbf{y}_B=F(\mathbf{B, D}),\quad\mathbf{y}_{\mathbf{C}_i}&=F(\mathbf{C}_i, \mathbf{D}), \\
i&=1,\cdots,p
\end{aligned}
\end{equation}
Here, $\mathbf{y}_A$, $\mathbf{y}_B$, and $\mathbf{y}_C$ will be acquired by executing the simulator.

The estimated first-order sensitivity indices and the total effect indices are defined as:
\begin{equation}
 S_i=\frac{Var[\mathbb{E}(Y|\eta_i)]}{Var(Y)}, \quad S_{T_i}=1-\frac{Var[\mathbb{E}(Y|\boldsymbol{\eta}_{\sim i})]}{Var(Y)}
\end{equation}
The variances can be calculated with the following equations proposed by Sobol \cite{SobolI} and Satelli \cite{SaltelliAndrea2002Mbuo}:
\begin{equation}
\begin{aligned}
 &Var[\mathbb{E}(Y|\eta_i)]=\frac{1}{M}\sum_{j=1}^{M}y_{A}^{(j)}y_{C_i}^{(j)}-(\mathbb{E}(Y))^2, \\\quad 
 &Var[\mathbb{E}(Y|\boldsymbol{\eta}_{\sim i})]=\frac{1}{M}\sum_{j=1}^{M}y_{B}^{(j)}y_{C_{i}}^{(j)}-(\mathbb{E}(Y))^{2}
\end{aligned}
\end{equation}
\begin{equation}
 Var(Y)=\frac{1}{M}\sum_{j=1}^M(y_A^{(j)})^2(\mathbb{E}(Y))^{2}
\end{equation}
The squared mean is approximated with the equation:
\begin{equation}
 (\mathbb{E}(Y))^{2}\approx\left(\frac{1}{M}\sum_{j=1}^My_A^{(j)}\right)\left(\frac{1}{M}\sum_{j=1}^My_B^{(j)}\right)\approx\left(\frac{1}{M}\sum_{j=1}^My_A^{(j)}\right)^2
\end{equation}

The perturbation method with a first-order Taylor approximation is applied for uncertainty propagation \cite{perturb}; that is, if the target positions have a known variance, the mean and variance of the simulator output can be quantified. The mean and variance of the input variables and the mean output are defined as:
\begin{equation}
 \mu_{\eta} = [\mu_{\eta_{1}}, \dots, \mu_{\eta_{p}}], ~\sigma^2_\eta = [\sigma^2_{\eta_1}, \dots, \sigma^2_{\eta_p}], ~y_0 = F(\mu_{\eta}, \mathbf{D})
\end{equation}
To find the propagated variance, the gradient of the function output with respect to $\eta$ is approximated with finite differences by giving a small increment $\delta$ to each $ \mu_{\eta_{i}}$, and the gradient is defined as:
\begin{equation}
 \frac{\partial y}{\partial \eta_i} \approx \frac{F(\mu_{\eta} + \delta \mathbf{e}_i, \mathbf{D}) -F(\mu_{\eta}, \mathbf{D}) }{\delta}
 \label{eq:grad_perturb}
\end{equation}
Where $\mathbf{e}_i$ is the canonical basis vector on the $i$-th dimension. Then, the variance of the output can be approximated as:
\begin{equation}
Var(y) \approx \sum^p_{i=1}\left (\frac{\partial y}{\partial \eta_i} \right )^2 \cdot \sigma^2_{\eta_i}
\end{equation}

The overall framework of the \ac{UQ} for the tether-net system is shown in Fig. \ref{fig:overall}. The target states and the \ac{MU} control parameters are sent to either the high-fidelity simulator or the surrogate-based environment to acquire the \acp{QoI}. For the Fixed-Control Net, the \ac{MU} control parameters are the same for all scenarios; in particular, each control parameter is selected to be the mean of the range defined for the corresponding variable in Table \ref{tab:optbounds}. The Active-Control Net will use the prior trained \ac{RL} policy to determine the control parameters. For the surrogate-based environment, the target states and \ac{MU} control parameters are sent to the \ac{LSTM}, and the output of the \ac{LSTM}, along with the target states, is used to calculate the input to the capture phase models to get the \acp{QoI}. The data is then utilized for the \ac{UQ} process mentioned in this section.

\begin{figure*}[htp!]
\centering
\includegraphics[width=1.0\linewidth]{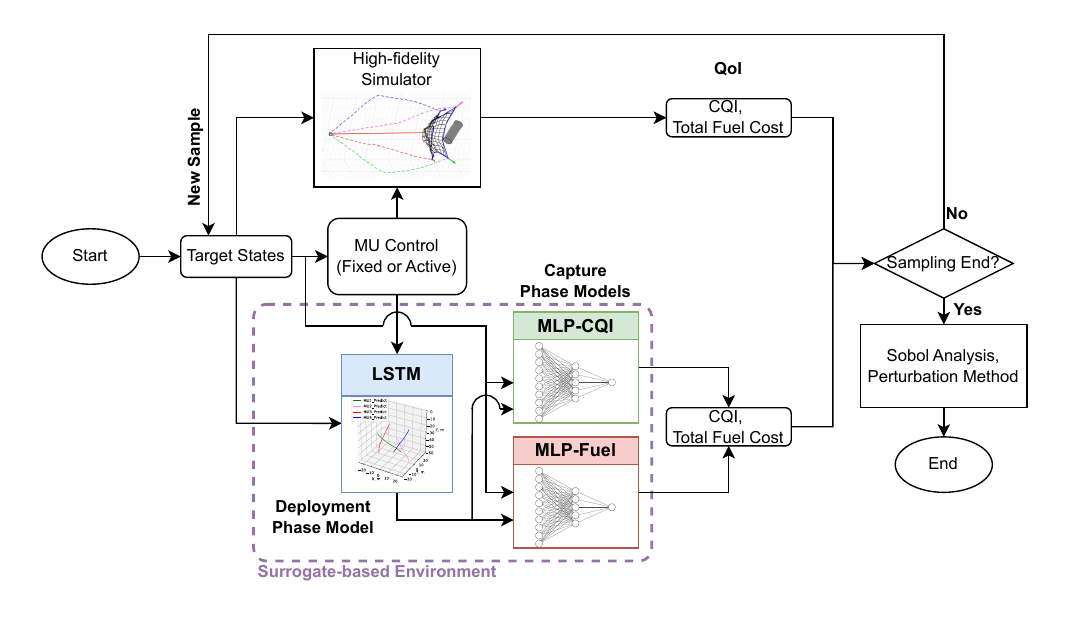}
\caption{Overall Framework of the UQ for Tether-Net System}
\label{fig:overall}
\end{figure*}

\section{Designs of Experiments for Uncertainty Analysis}\label{sec:case_study}

The sensitivity analysis provides a broad understanding of how the tether-net system performs across the entire input space. In contrast, the analysis employing the perturbation method is applied using a known variance derived from parallel research to evaluate whether the system can accurately capture and respond to small input perturbations. In addition, a sampling-based method was used along with the perturbation method to serve as a reference for evaluating the performance of the latter. This comparison helps assess whether the perturbation approach can effectively approximate the response of the system under small variations in the input space. The design of the experiments is based on the following motivations:
\begin{itemize}
 \item Investigate how uncertainty in the target’s initial X, Y, and Z coordinates influences the \acp{QoI} for a Fixed-Control Net system, and analyze how this uncertainty propagates through the simulation.
 
 \item Examine the impact of target position uncertainty on the QoI in an Active-Control Net system, and assess the corresponding uncertainty propagation.
 
 \item Evaluate the effects on sensitivity and uncertainty propagation when the high-fidelity simulator is replaced with a combination of surrogate models.
\end{itemize}

\noindent Therefore, the following case studies have been conducted.
\begin{itemize}
 \item \textbf{Fixed-Control Net \ac{UQ}:} Performing \ac{UQ} on the Fixed-Control Net configuration is essential to establish a baseline understanding of the inherent behavior of the tether-net system in the absence of active control. This analysis provides valuable insights into how variations in initial conditions impact the fuel consumption and final \ac{CQI}. For the sensitivity analysis, the sample number $M$ is set to 500, and the variable number $p$ is 3; therefore, the total number of samples for matrix $\mathbf{A}$, $\mathbf{B}$, and $\mathbf{C}$ is 2,500. The value of vector $\mathbf{D}$ is set to the mean of the bounds. For the perturbation method, the variance $\sigma^2_{\eta_i}$ is set to be $2.77^2$ m$^2$ for all ${\eta_i}$.
\item \textbf{Active-Control Net \ac{UQ}:} The same tasks will be conducted as the previous case study, but the \ac{MU} control vector $\mathbf{X}$ is determined by the \ac{RL} policy. Here, we assumed the perception of the tether-net system is perfect, so the controls determined by the \ac{RL} are based on precise target positions as inputs to the \ac{RL} policy. In other words, the \ac{RL} policy is considered as part of the function evaluations to get the QoIs. The motivation of this case study is to explore the scenarios where the initial launching position of the net is noised.
 \item \textbf{Active-Control Net \ac{UQ} with Perception Uncertainty:} In reality, perception is imperfect, leading to noisy estimation of the state of the target debris. Thus, in this case study, we conducted the sensitivity analysis with the target position coordinates defined to be at the mean of the considered range, but the inputs to the policy are sampled from the range of $\mathcal{U}[-4.8, +4.8]$ m. 
 This case study gives a basic understanding of how robust the \ac{RL}-aided control is in this particular scenario, and quantifies the uncertainty caused by this noisy observation.
 \item \textbf{\ac{UQ} on the Surrogate-Model-Based Environment:} The same tasks for both Fixed-Control and Active-Control Net systems are re-evaluated with the surrogate models in place of the high-fidelity simulator, but the sampling number $M$ has been set to 4,000. Additionally, a sensitivity analysis for all input variables (i.e., the target states and \ac{MU} control parameters) is performed, and the range of each input is defined in Table \ref{tab:optbounds}.
\end{itemize}

\section{Results and Discussion}\label{sec:results}

Here we present the results of the case studies introduced in the previous section. The sampling part for the sensitivity analysis was executed on a group of Intel Xeon Gold 6230 with 187 GB RAM for efficiency, while the perturbation method and the surrogate-model-related computing were executed on an AMD Ryzen 9 5950X-16-Core processor, 64 GB RAM, and 12 GB NVIDIA GeForce RTX-3060 GPU. Note that, due to estimator variance and finite sample effects, especially when dealing with complex or noisy black-box models, it is not unusual for computed first-order or total-order Sobol indices (i.e., $S_i$ and $S_{T_i}$) to appear slightly negative. This could happen even though their true theoretical values are non-negative by definition \cite{SobolI, SaltelliA2008GSAT}. 
Therefore, to maintain interpretability and adhere to the physical meaning of these indices as measures of variance contribution, it is standard to clip or truncate negative estimates to zero in post-processing; this practice ensures consistency in the sensitivity interpretation without significantly biasing the overall analysis. Accordingly, in the following presentation of results, the negative Sobol indices have been truncated to zeros. The results of the Sobol indices for the previously mentioned case studies are listed in Table \ref{tab:sobol_collect} and \ref{tab:sobol_collect_ALL}. The details of the results for each case study are discussed in the upcoming paragraphs.
\begin{table*}[ht]
\centering
\footnotesize
\caption{Sobol Indices of the High Fidelity Case Studies}
\label{tab:sobol_collect}
\begin{tabular}{|c|ccc|ccc|ccc|}
\hline
\textbf{Sobol Indices} 
& \multicolumn{3}{c|}{\textbf{Fixed-Control}} 
& \multicolumn{3}{c|}{\textbf{Active-Control}} 
& \multicolumn{3}{c|}{\makecell{\textbf{Active-Control} \\ \textbf{with Perception} \\ \textbf{Uncertainty}}} \\
\hline
& $X_D$ & $Y_D$ & $Z_D$ & $X_D$ & $Y_D$ & $Z_D$ & $X_D$ & $Y_D$ & $Z_D$ \\
\hline
CQI $S_i$      & 0.00 & 0.01 & 0.00 & 0.09 & 0.06 & 0.00 & 0.16 & 0.12 & 0.06 \\
CQI $S_{T_i}$  & 0.88 & 0.88 & 0.46 & 1.00 & 0.86 & 0.93 & 0.86 & 0.85 & 0.38 \\
\hline
Tot. Fuel Cost $S_i$      & 0.21 & 0.00 & 0.07 & 0.09 & 0.05 & 0.76 & 0.01 & 0.00 & 1.03 \\
Tot. Fuel Cost $S_{T_i}$  & 0.92 & 0.94 & 0.36 & 0.34 & 0.24 & 0.88 & 0.02 & 0.08 & 0.99 \\
\hline
\end{tabular}
\end{table*}

\begin{table*}[ht]
\centering
\footnotesize
\setlength{\tabcolsep}{3pt} 
\caption{Sobol Indices of the Surrogate-based Environment on All the Inputs}
\label{tab:sobol_collect_ALL}
\begin{tabular}{|c|ccc|ccc|ccc|ccccccccc|}
\hline
\textbf{Sobol Indices} 
& \multicolumn{3}{c|}{\makecell{\textbf{Target} \\ \textbf{Position}}}
& \multicolumn{3}{c|}{\makecell{\textbf{Target} \\ \textbf{Orientation}}} 
& \multicolumn{3}{c|}{\makecell{\textbf{Target} \\ \textbf{Angular Velocity}}}
& \multicolumn{9}{c|}{\makecell{\textbf{MU Control}}} \\
\hline
& $X_D$ & $Y_D$ & $Z_D$ & $\omega_X$ & $\omega_Y$ & $\omega_Z$ & $O_X$ & $O_Y$ & $O_Z$ & $\psi_{T,1}$ &  $\psi_{T,2}$  & $\psi_{T,3}$ & $\psi_{T,4}$ & $\theta_{T,1}$ & $\theta_{T,2}$ & $\theta_{T,3}$ & $\theta_{T,4}$ &  $F_T$\\
\hline
CQI $S_i$      & 0.00 & 0.00 & 0.06 & 0.00 & 0.00 & 0.00 & 0.00 & 0.00 & 0.00 & 0.00 & 0.00 & 0.00 & 0.00 & 0.00 & 0.00 & 0.00 & 0.00 & 0.08\\
CQI $S_{T_i}$      & 0.29 & 0.33 & 0.38 & 0.07 & 0.07 & 0.08 & 0.06 & 0.06 & 0.06 & 0.46 & 0.48 & 0.46 & 0.49 & 0.10 & 0.09 & 0.07 & 0.06 & 0.26\\
\hline
Tot. Fuel Cost $S_i$  & 0.00 & 0.00 & 0.00 & 0.00 & 0.00 & 0.00 & 0.00 & 0.00 & 0.00 & 0.01 & 0.00 & 0.00 & 0.00 & 0.00 & 0.00 & 0.00 & 0.00 & 0.00\\
Tot. Fuel Cost $S_{T,i}$  & 0.10 & 0.16 & 0.38 & 0.14 & 0.12 & 0.11 & 0.14 & 0.14 & 0.11 & 0.35 & 0.42 & 0.40 & 0.42 & 0.70 & 0.76 & 0.66 & 0.68 & 0.26\\
\hline
\end{tabular}
\end{table*}

\subsection{Fixed-Control Net UQ}
\begin{figure*}[htp!]
 \centering
 \begin{subfigure}[t]{0.48\linewidth}
 \centering
 \includegraphics[width=\linewidth]{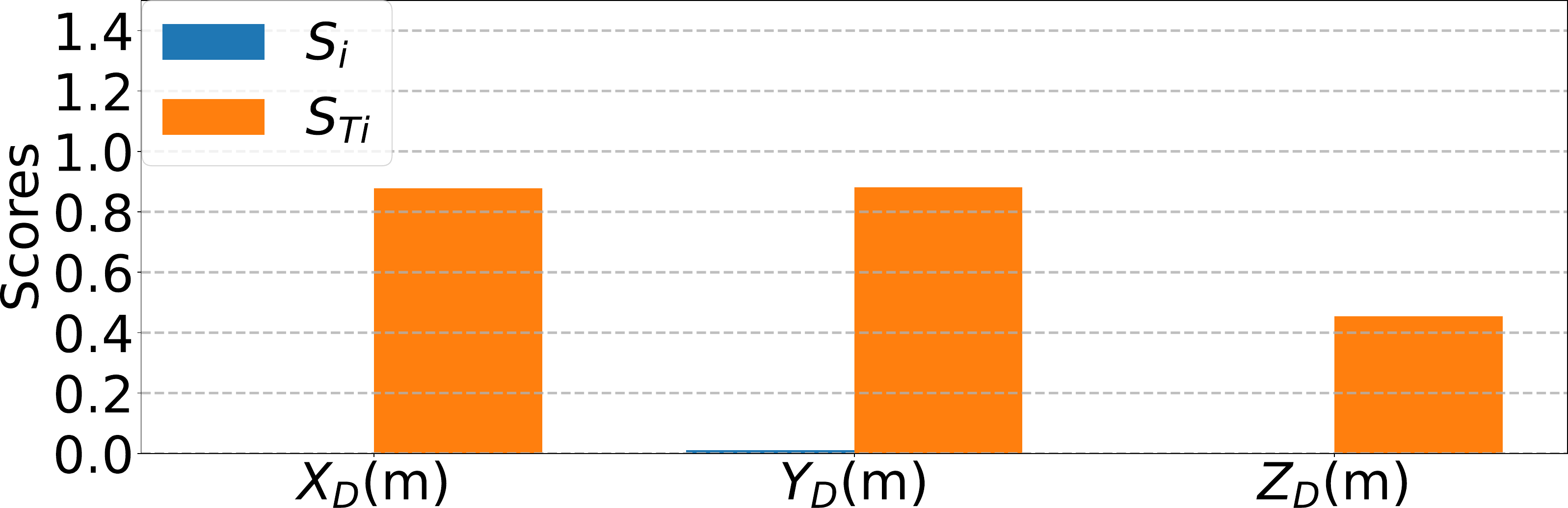}
 \caption{Sobol Indices of Target State (CQI)}
 \label{fig:sobol-Fixed-Control}
 \end{subfigure}
 \hfill
 \begin{subfigure}[t]{0.48\linewidth}
 \centering
 \includegraphics[width=\linewidth]{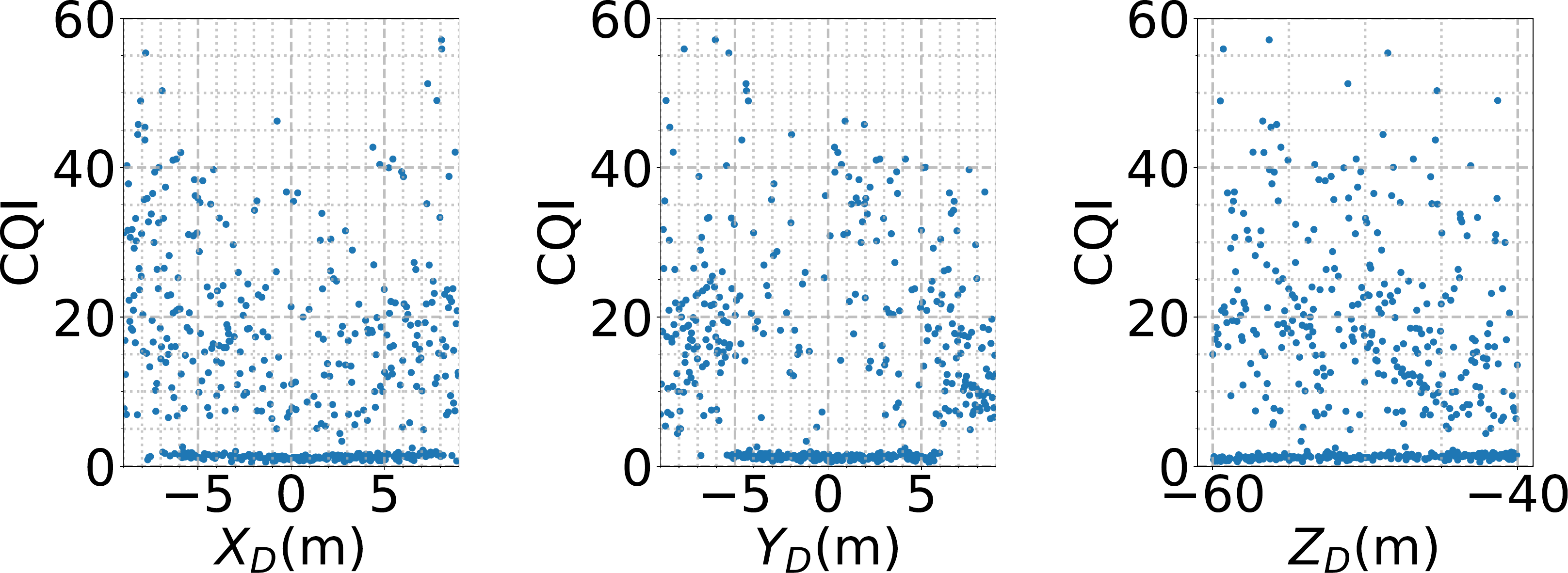}
 \caption{Scatterplot of Fixed-Control Net CQI}
 \label{fig:scatter-Fixed-Control}
 \end{subfigure}
 
 \vspace{1em} 

 \begin{subfigure}[t]{0.48\linewidth}
 \centering
 \includegraphics[width=\linewidth]{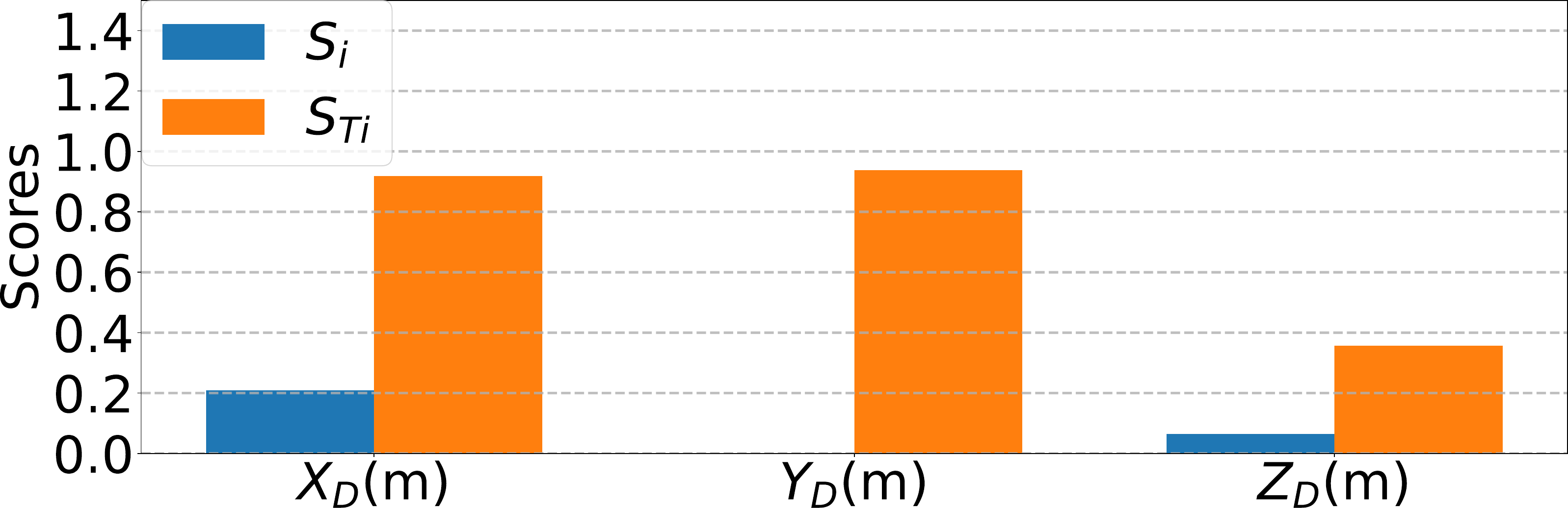}
 \caption{Sobol Indices of Target State (Total Fuel)}
 \label{fig:sobol-Fixed-Control-fuel}
 \end{subfigure}
 \hfill
 \begin{subfigure}[t]{0.48\linewidth}
 \centering
 \includegraphics[width=\linewidth]{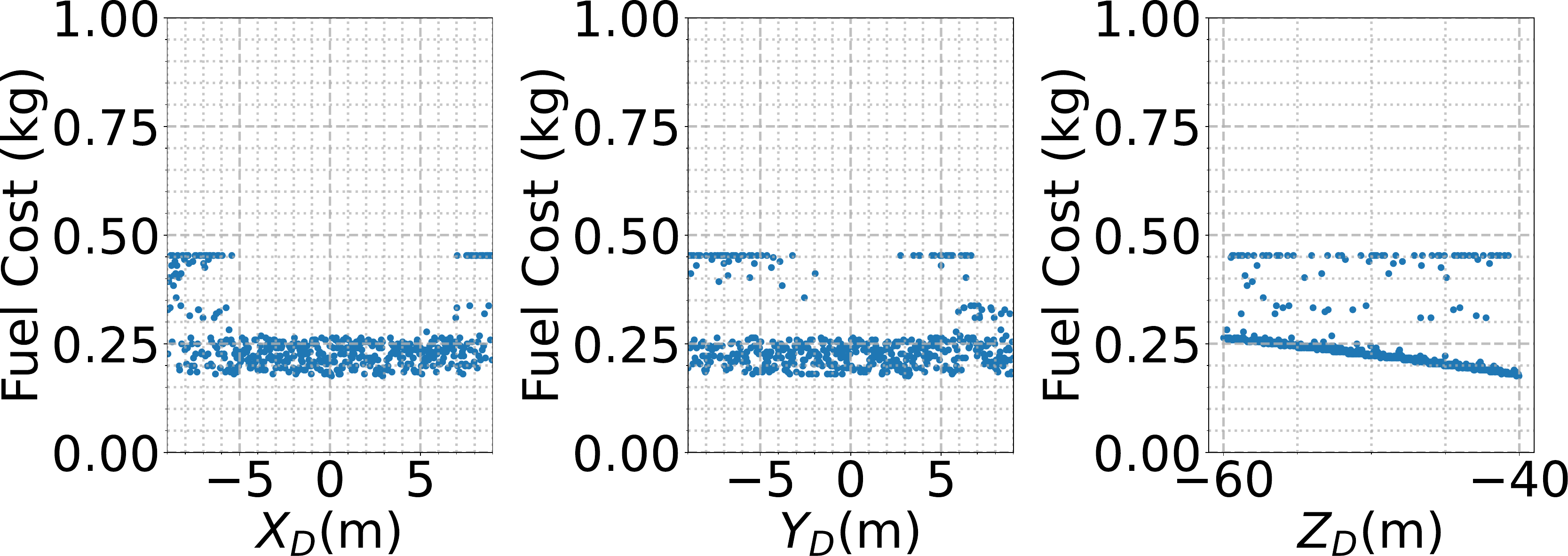}
 \caption{Scatterplot of Fixed-Control Net Total Fuel}
 \label{fig:scatter-Fixed-Control-fuel}
 \end{subfigure}

 \vspace{1em} 
 
 \begin{subfigure}[t]{0.48\linewidth}
 \centering
 \includegraphics[width=\linewidth]{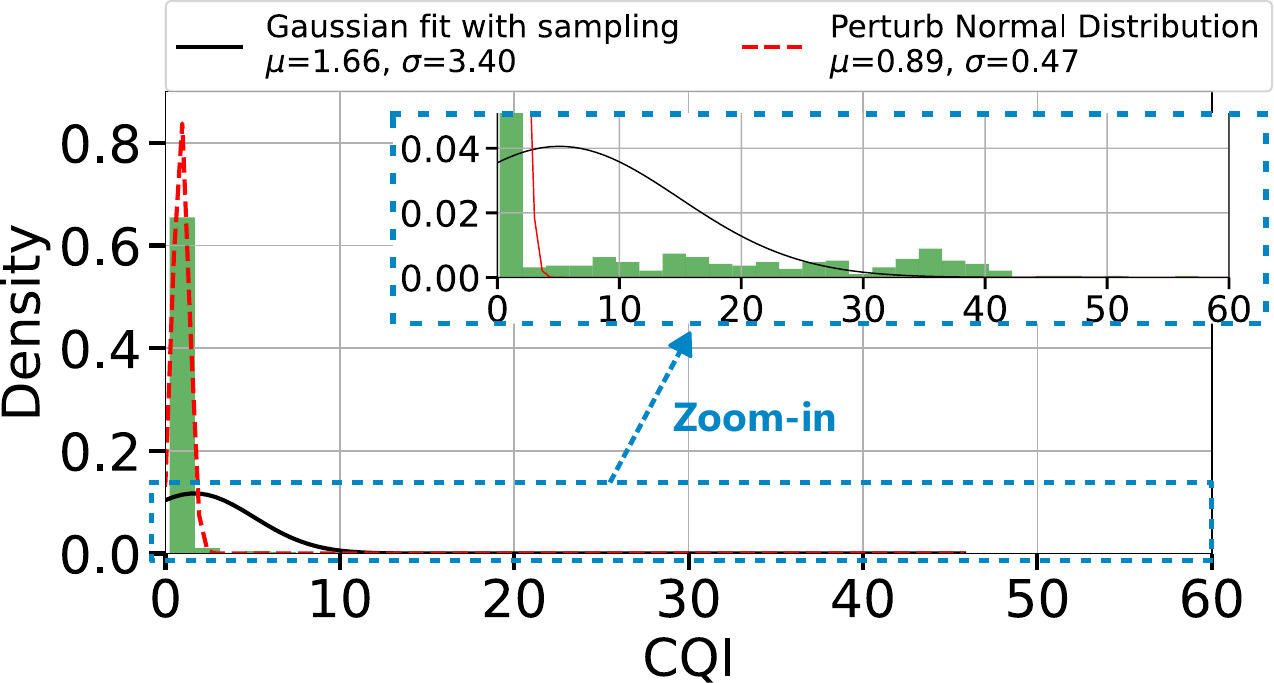}
 \caption{Perturbation Result for CQI}
 \label{fig:perturb_passive_cqi}
 \end{subfigure}
 \hfill
 \begin{subfigure}[t]{0.48\linewidth}
 \centering
 \includegraphics[width=\linewidth]{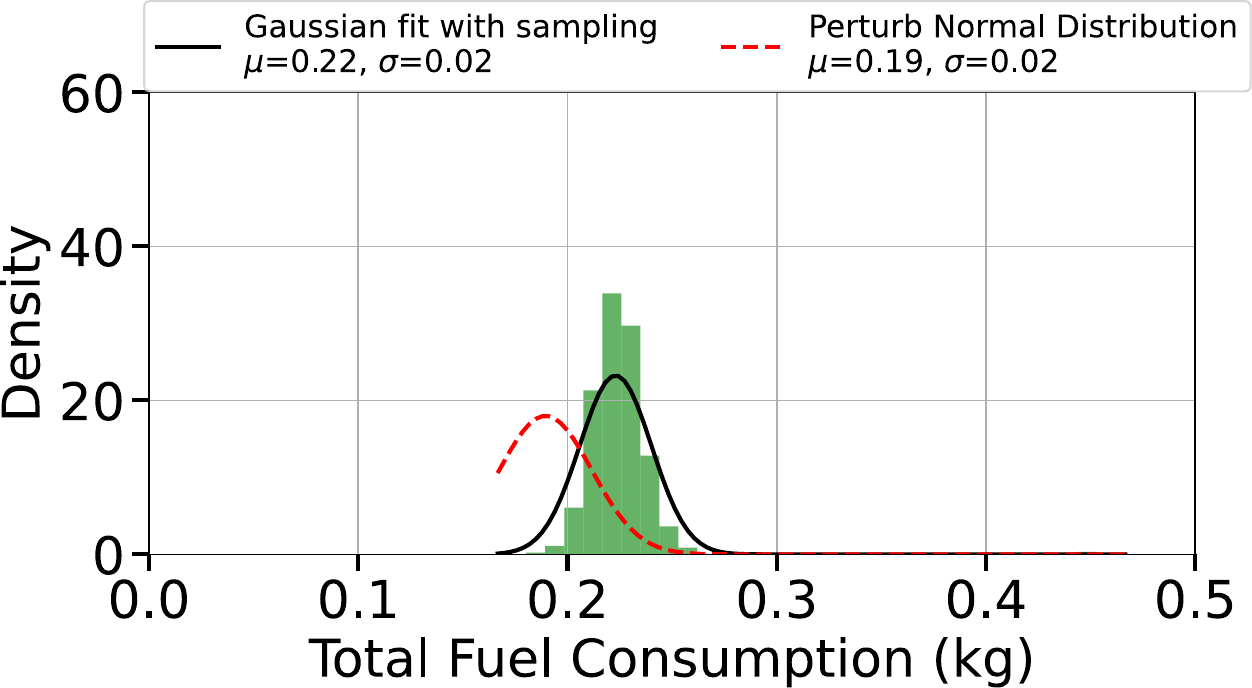}
 \caption{Perturbation Result for Total Fuel Consumption}
 \label{fig:perturb_passive_fuel}
 \end{subfigure}

 \caption{Comparison of Sobol Indices and Scatterplots for Fixed-Control Net on Two Metrics: CQI and Total Fuel}
 \label{fig:HF-passive}
\end{figure*}


For the \ac{UQ} of the Fixed-Control Net, Fig. \ref{fig:sobol-Fixed-Control} shows that the first-order Sobol indices for \ac{CQI} about $X_D$ and $Z_D$ are truncated to 0, and $Y_D$ is close to 0, which is also indicated by Fig. \ref{fig:scatter-Fixed-Control} since no clear trend can be noticed. However, the high $S_{T_i}$ value indicates that when collaborating with other variables, $X_D$ and $Y_D$ are more influential than $Z_D$. This is to be expected,  as displacements in the X- and Y-directions make the capture more difficult and can lead to a failed wrapping, resulting in highly varying \ac{CQI} values. This also holds for the fuel consumption, as shown in Figs. \ref{fig:sobol-Fixed-Control-fuel} and \ref{fig:scatter-Fixed-Control-fuel}. Moreover, compared with the plot of the \ac{CQI}, it can be seen that when the X- and Y-coordinates of the target position are close to the mean value, the fuel consumption is significantly lower. The fuel consumption increases after the displacements exceed around $\pm$5 m for both X and Y, which is reasonable because beyond this threshold, the central part of the net is likely to miss the target, resulting in delayed thrust deactivation for the system. The scatter plot of $Z_D$ shows that when the target is closer, the fuel consumption is lower; however, $S_i$ is not large because the value has not converged, and many samples are still scattered between the region above 0.3 kg. Note that since there is an upper bound for the total thrusting time, there is a corresponding upper bound of the fuel consumption (see Section \ref{sec:sim_mod}). For this work, if a capture ends up close to the upper limit for the fuel consumption, it represents a situation where the net missed the target; this explains why there are values concentrated close to the top of the fuel consumption plot. It should also be noted that a \ac{CQI} value smaller than 2.5 indicates that a successful capture \cite{liu2023learning}; this explains why there are values concentrated at the bottom of the \ac{CQI} plot. Since the scatter plot is an illustration of $S_i$, this explains why $X_D$ and $Y_D$ have higher $S_i$ values than $Z_D$.

Figures \ref{fig:perturb_passive_cqi} and \ref{fig:perturb_passive_fuel} depict the \ac{CQI} and the total fuel consumption, respectively, for the perturbation method with the Fixed-Control Net. Only 4 simulations were needed to obtain the variance and mean of the output when each coordinate of the target position has a $2.77^2$ m$^2$ variance. Ideally, the value of $\delta$ should be small in Eq. \eqref{eq:grad_perturb} for approximating the gradient, but because the simulation frequency is not low enough to capture the change in fuel consumption when the target position varies less than a scale of 0.1 m, we used $\delta=1$ for the calculation. The plot shows that the perturbation method's variance and mean are close to the result of 1,000 Monte-Carlo Sampling for the fuel consumption. For the \ac{CQI}, the mean and variance of the perturbation result are a lot different from the Monte-Carlo result; this mismatch is because the sampling result is highly right-skewed (as the zoomed-in plot is showing). Though the perturbation method does not work well for the \ac{CQI}, it converged on the highest-density samples.

\subsection{Active-Control Net UQ}
\begin{figure*}[htp!]
 \centering
 \begin{subfigure}[t]{0.48\linewidth}
 \centering
 \includegraphics[width=\linewidth]{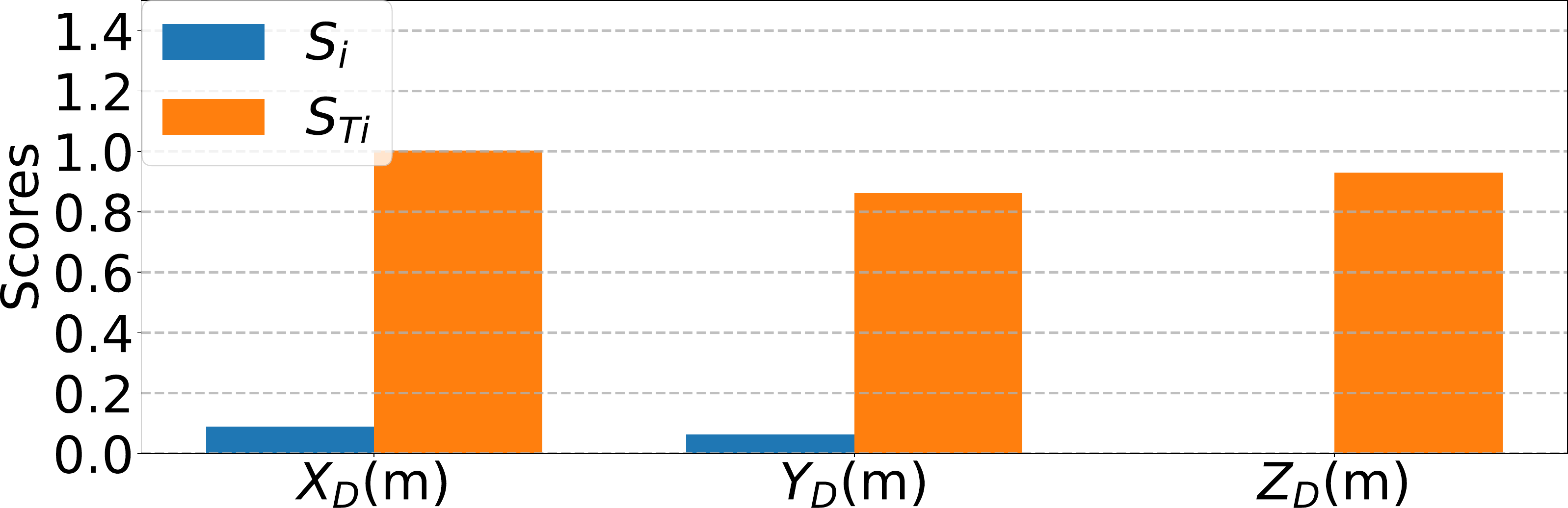}
 \caption{Sobol Indices of Target State (CQI)}
 \label{fig:sobol-active}
 \end{subfigure}
 \hfill
 \begin{subfigure}[t]{0.48\linewidth}
 \centering
 \includegraphics[width=\linewidth]{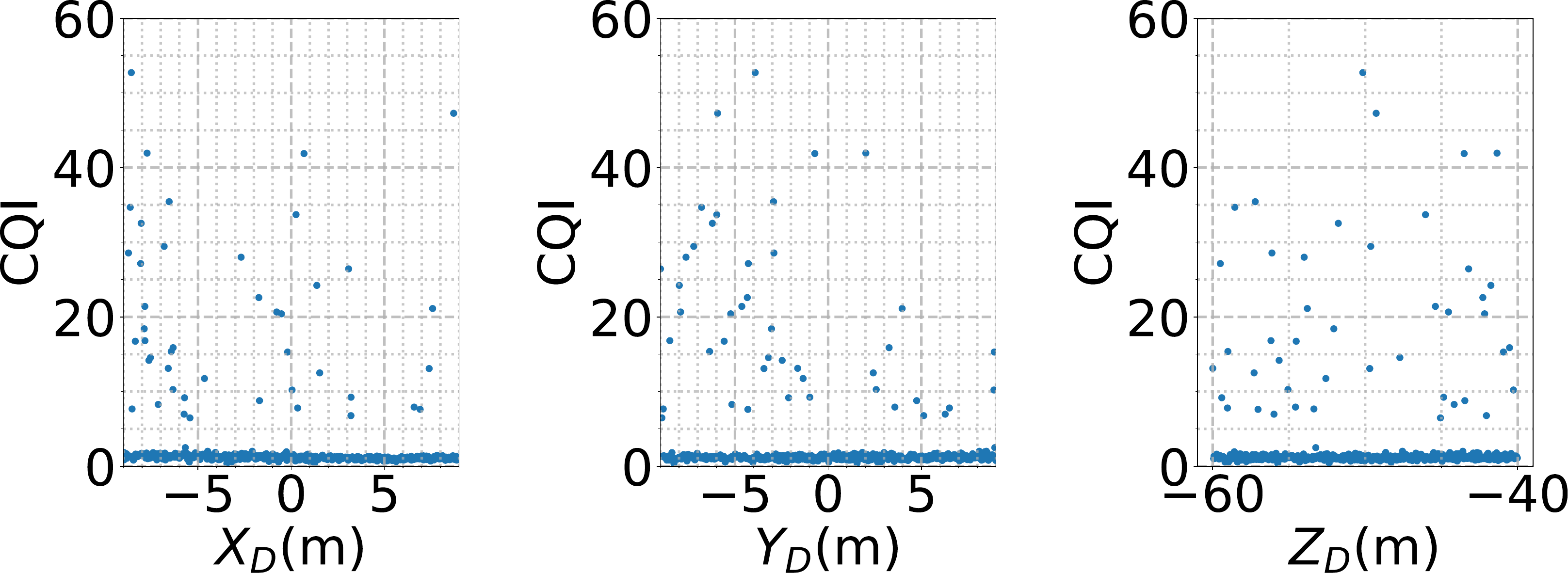}
 \caption{Scatterplot of Active-Control Net CQI}
 \label{fig:scatter-active}
 \end{subfigure}

 \vspace{1em} 

 \begin{subfigure}[t]{0.48\linewidth}
 \centering
 \includegraphics[width=\linewidth]{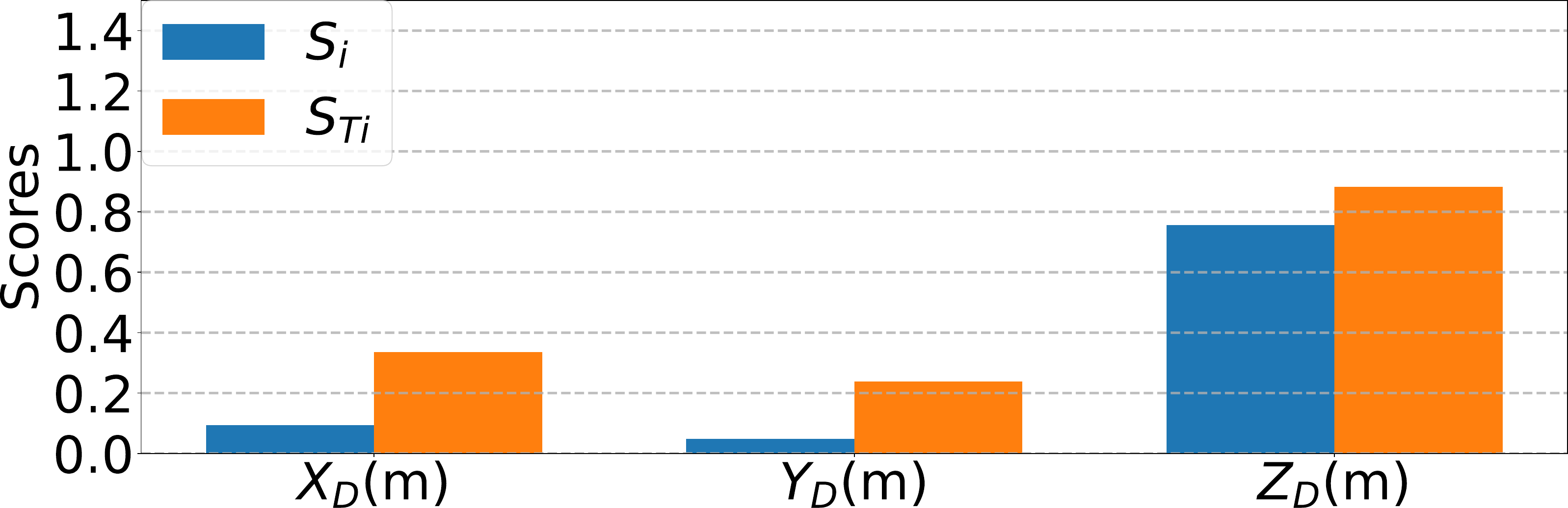}
 \caption{Sobol Indices of Target State (Total Fuel)}
 \label{fig:sobol-active-fuel}
 \end{subfigure}
 \hfill
 \begin{subfigure}[t]{0.48\linewidth}
 \centering
 \includegraphics[width=\linewidth]{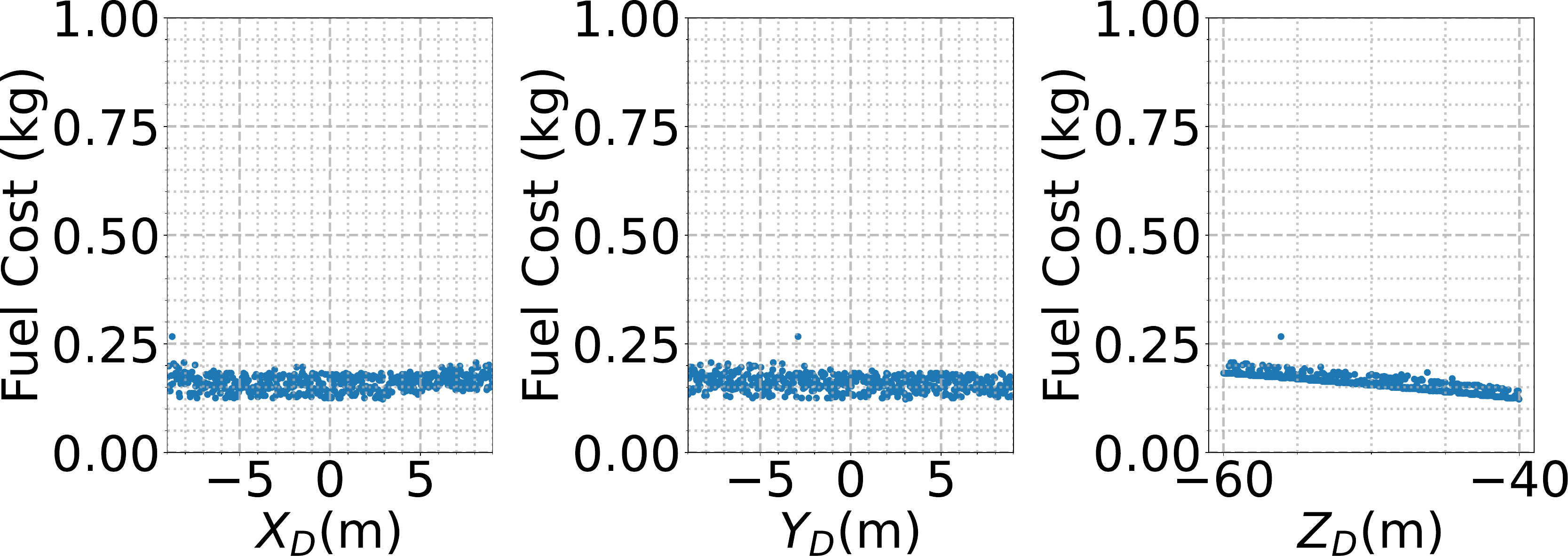}
 \caption{Scatterplot of Active-Control Net Total Fuel}
 \label{fig:scatter-active-fuel}
 \end{subfigure}

 \vspace{1em}
 \begin{subfigure}[t]{0.48\linewidth}
 \centering
 \includegraphics[width=\linewidth]{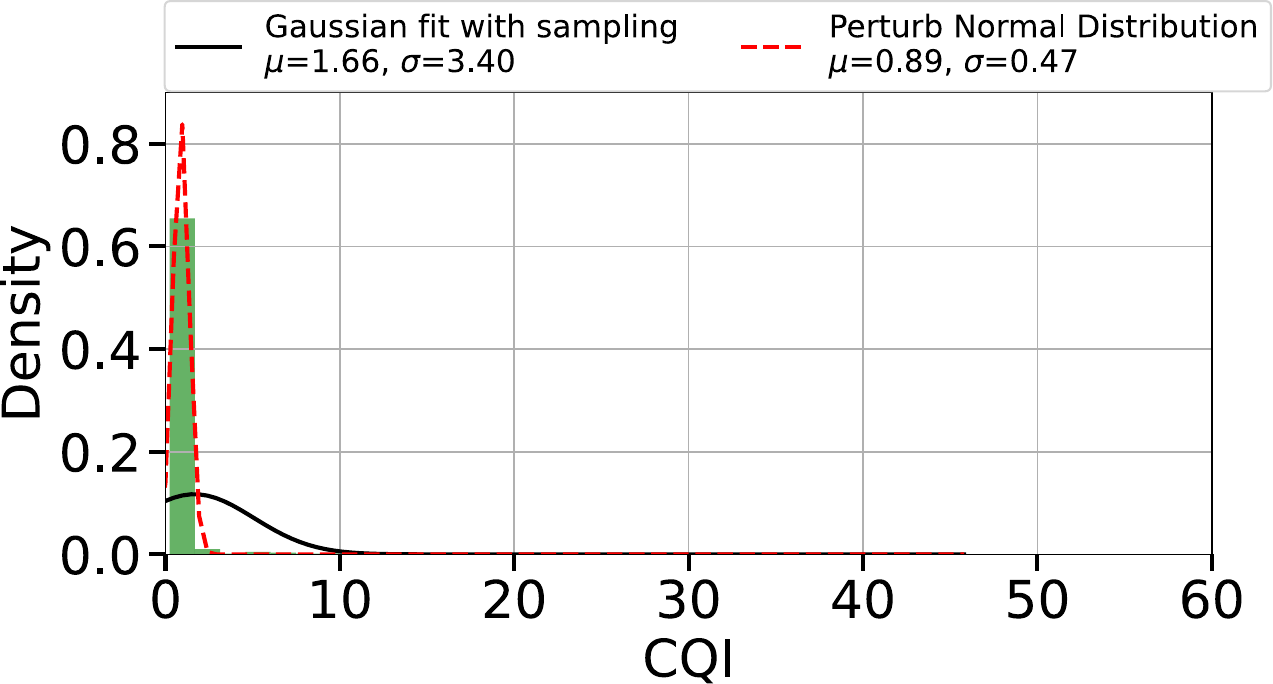}
 \caption{Perturbation Result for CQI}
 \label{fig:perturb_active_cqi}
 \end{subfigure}
 \hfill
 \begin{subfigure}[t]{0.48\linewidth}
 \centering
 \includegraphics[width=\linewidth]{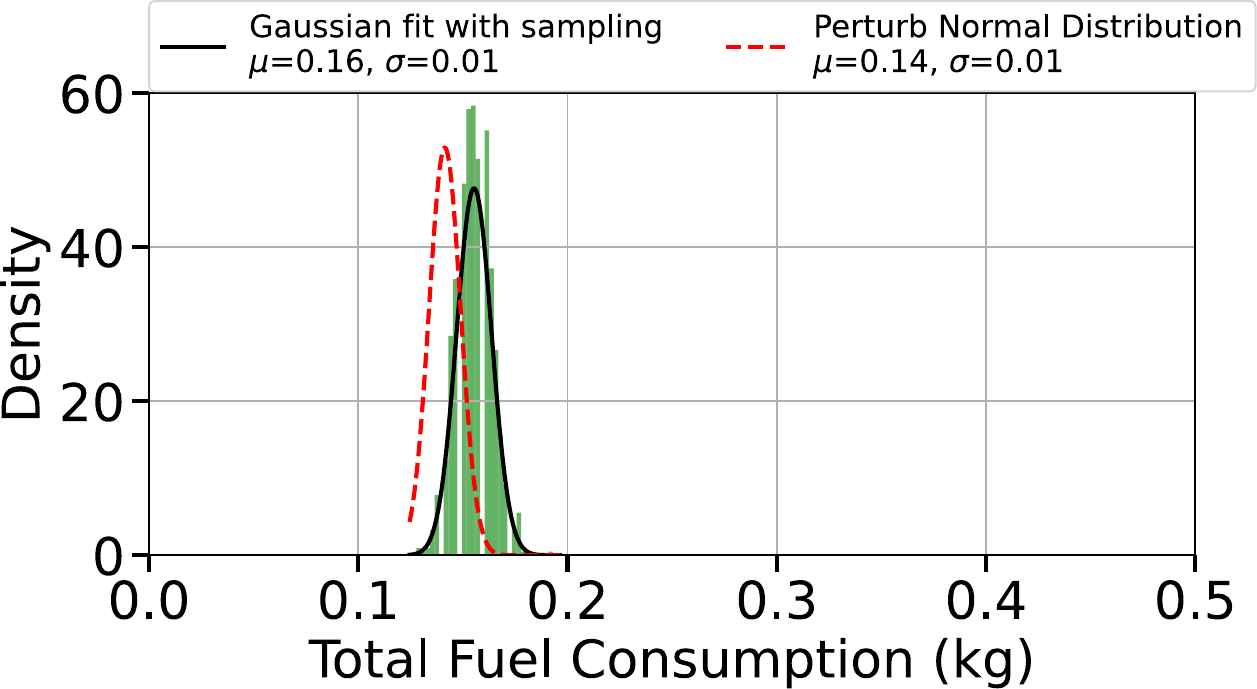}
 \caption{Perturbation Result for Total Fuel Consumption}
 \label{fig:perturb_active_fuel}
 \end{subfigure}

 \caption{Comparison of Sobol Indices and Scatterplots for Active-Control Net on Two Metrics: CQI and Total Fuel}
 \label{fig:HF-active}
\end{figure*}

Figure \ref{fig:HF-active} shows the result of the sensitivity analysis and the perturbation method result of the Active-Control Net. The target states are the same as the ones in the previous case study, but the control inputs are produced by the \ac{RL} policy instead. It can be observed that in the scatter plots for both the \ac{CQI} and the total fuel consumption, the values are much less sparse. This proves that the \ac{RL} policy we pre-trained has a high successful capture rate. In the 2,500 samples collected for the sensitivity analysis, the Active-Control Net achieved 87.64\% successful capture rate; thus, the \ac{CQI} and the fuel consumption are less influenced by failed captures. Figures \ref{fig:sobol-active} and \ref{fig:scatter-active} show that $X_D$, $Y_D$ and $Z_D$ all have similarly low $S_i$ and high $S_{T_i}$: this is because the majority of the captures are successful, resulting in low \ac{CQI} values. The fuel consumption results of Figs. \ref{fig:sobol-active-fuel} and \ref{fig:scatter-active-fuel} show that $Z_D$ is a lot more influential than $X_D$ and $Y_D$, in addition to having a strong independent effect. The scatter plots show there is a clear linear relationship for $Z_D$. The larger range of $Z_D$ considered -- compared to $X_D$ and $Y_D$ -- could also be the cause of this higher sensitivity. Another factor that causes $Z_D$ to be more influential is that the activation of the closing mechanism depends on Z components of the target-\acp{MU} relative positions. Therefore, because of these factors, the fuel consumption of the tether-net system is mainly affected by $Z_D$. 
In Figs. \ref{fig:perturb_active_cqi} and \ref{fig:perturb_active_fuel}, the perturbation method results also show mean and variance values closer to the Monte-Carlo sampling result. This indicates that the perturbation method can be a promising approach to estimate the propagation of the uncertainty for this tether-net system -- in contrast to the Fixed-Control Net -- since only four function evaluations are sufficient. 

\subsection{Active-Control Net UQ with Perception Uncertainty:}
\begin{figure*}[htp!]
 \centering
 \begin{subfigure}[t]{0.48\linewidth}
 \centering
 \includegraphics[width=\linewidth]{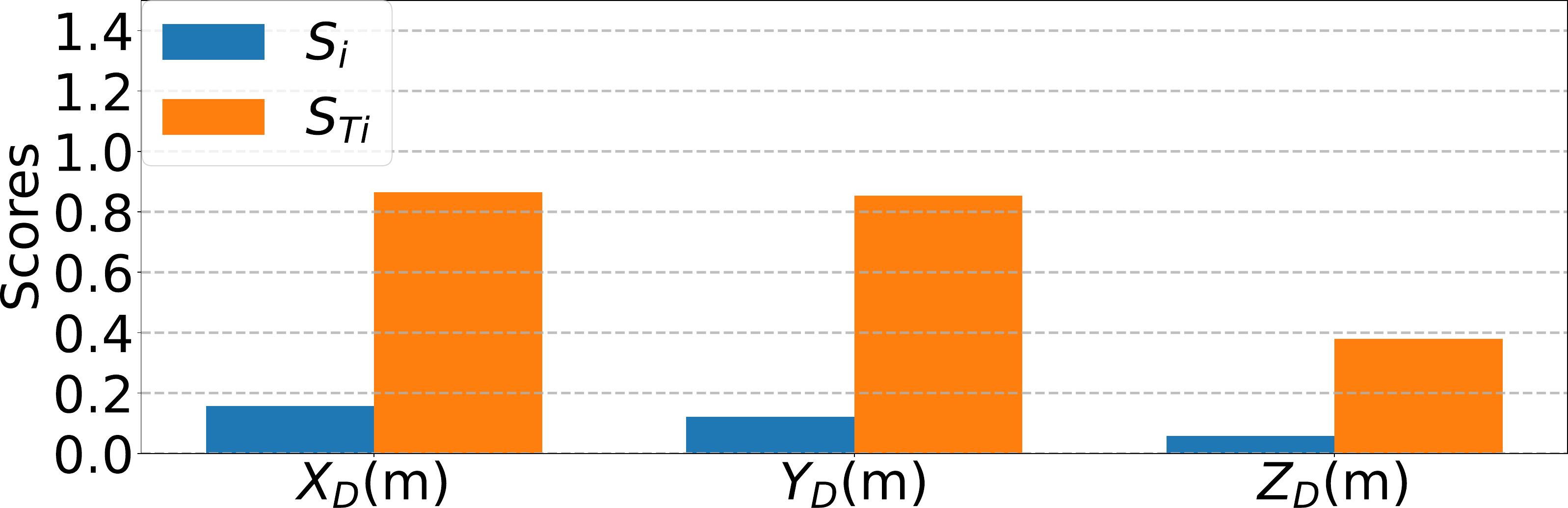}
 \caption{Sobol Indices of Target State (CQI)}
 \label{fig:sobol-active-percept}
 \end{subfigure}
 \hfill
 \begin{subfigure}[t]{0.48\linewidth}
 \centering
 \includegraphics[width=\linewidth]{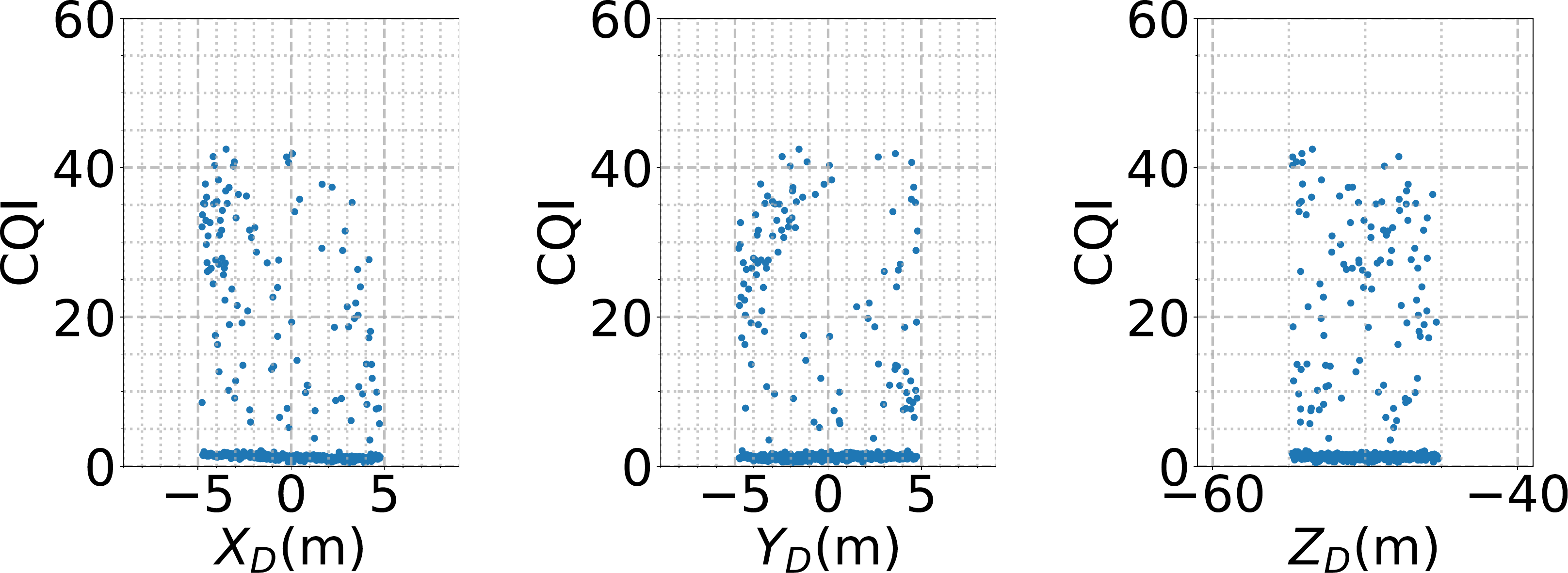}
 \caption{Scatterplot of Active-Control Net CQI}
 \label{fig:scatter-active-percept}
 \end{subfigure}

 \vspace{1em} 

 \begin{subfigure}[t]{0.48\linewidth}
 \centering
 \includegraphics[width=\linewidth]{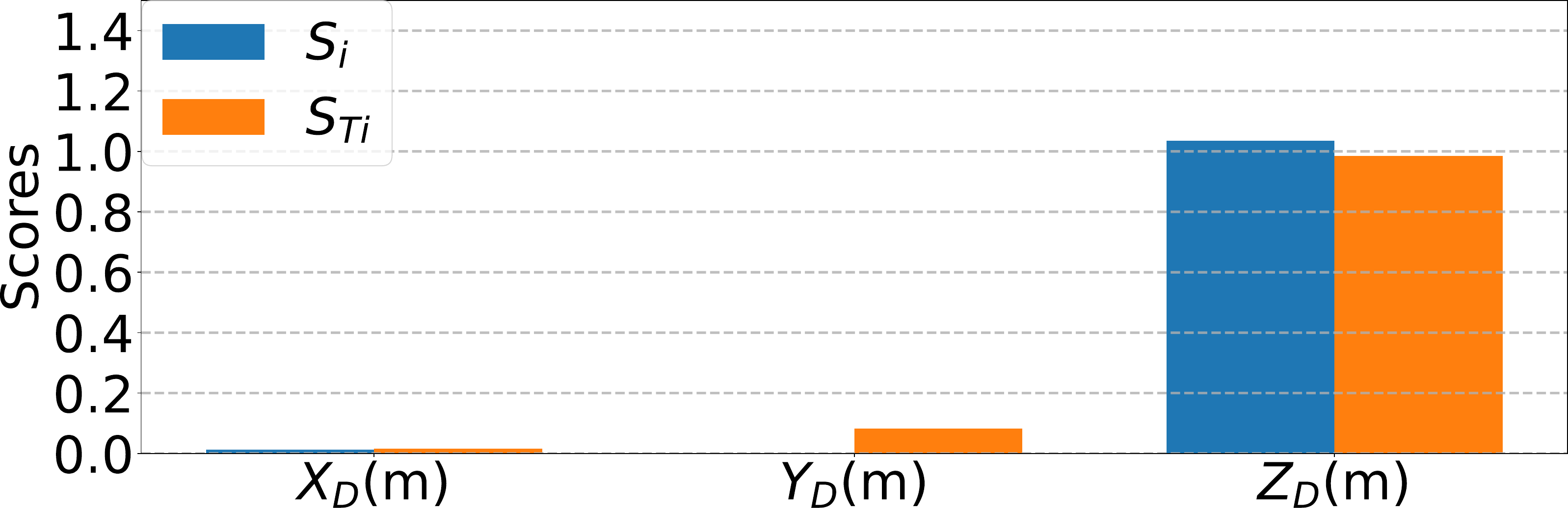}
 \caption{Sobol Indices of Target State (Total Fuel)}
 \label{fig:sobol-active-fuel-percept}
 \end{subfigure}
 \hfill
 \begin{subfigure}[t]{0.48\linewidth}
 \centering
 \includegraphics[width=\linewidth]{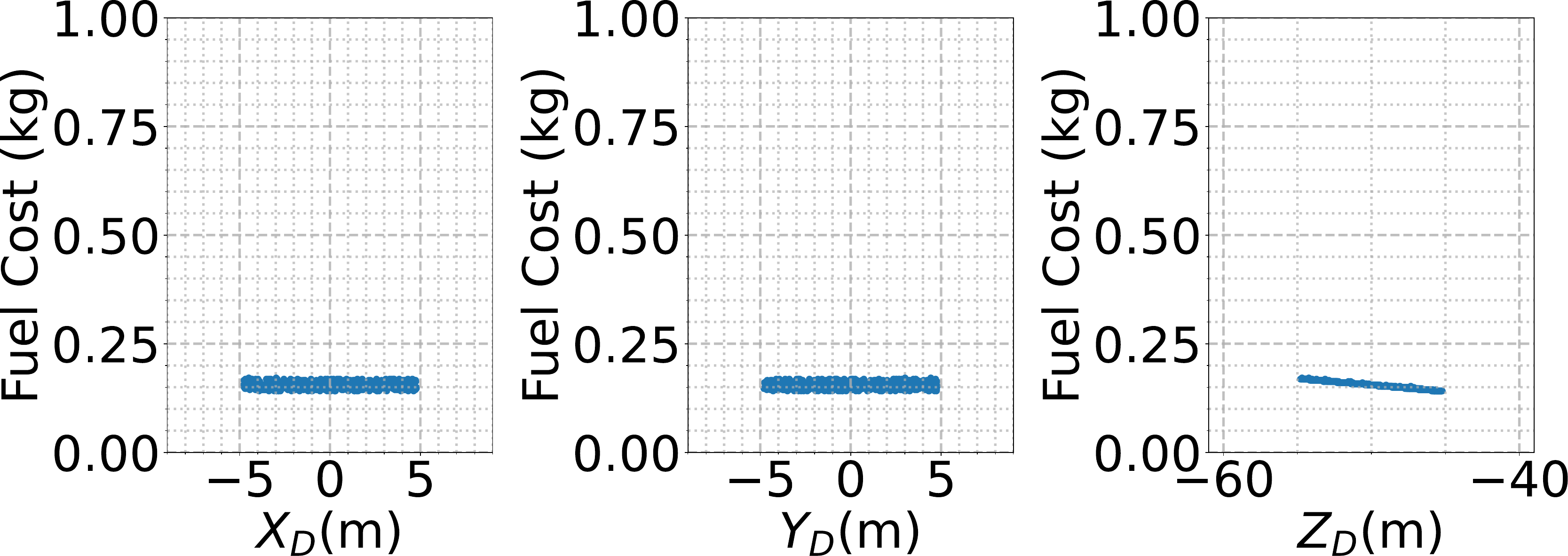}
 \caption{Scatterplot of Active-Control Net Total Fuel}
 \label{fig:scatter-active-fuel-percept}
 \end{subfigure}

 \caption{Comparison of Sobol Indices and Scatterplots for Active-Control Net with Perception Uncertainty on Two Metrics: CQI and Total Fuel}
 \label{fig:HF-active-percept}
\end{figure*}

Figure \ref{fig:HF-active-percept} shows the Active-Control Net sensitivity analysis when the inputs -- or the \textit{observations} -- for the \ac{RL} policy are noised; this could be induced by sensor inaccuracies or perception estimation errors in real missions. In this scenario, the actual target states are fixed, but the controls of the net are different due to the noisy observation. For the \ac{CQI}, the noises for the observed $X_D$ and $Y_D$ are more influential; this is reasonable because the noise on these two axes encourages the \ac{RL} policy to produce the control that may lead to missed capture more than noise on the Z direction. This also explains why in the fuel consumption analysis in Figs. \ref{fig:sobol-active-fuel-percept} and \ref{fig:scatter-active-fuel-percept}, only $Z_D$ has a noticeable impact.



\subsection{UQ on the Surrogate-Model-Based Environment}

\begin{figure*}[htp!]
 \centering
 \begin{subfigure}[t]{0.48\linewidth}
 \centering
 \includegraphics[width=\linewidth]{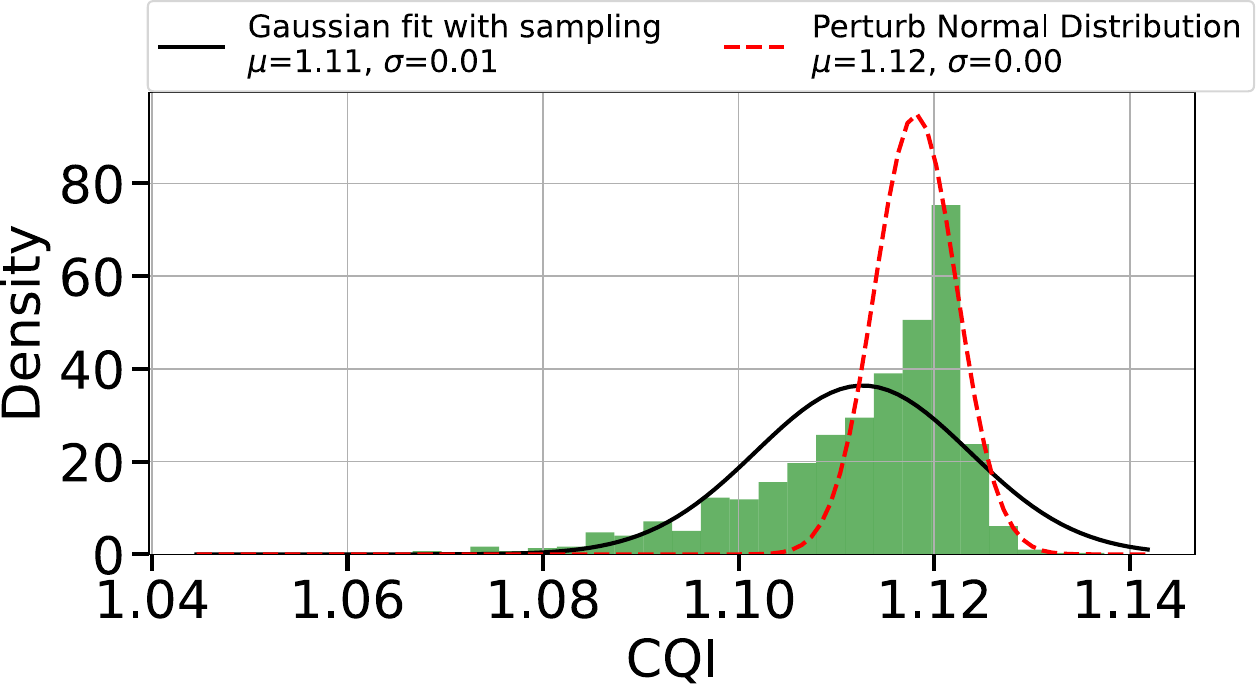}
 \caption{Perturbation Result for CQI}
 \label{fig:perturb_active_cqi_SU}
 \end{subfigure}
 \hfill
 \begin{subfigure}[t]{0.48\linewidth}
 \centering
 \includegraphics[width=\linewidth]{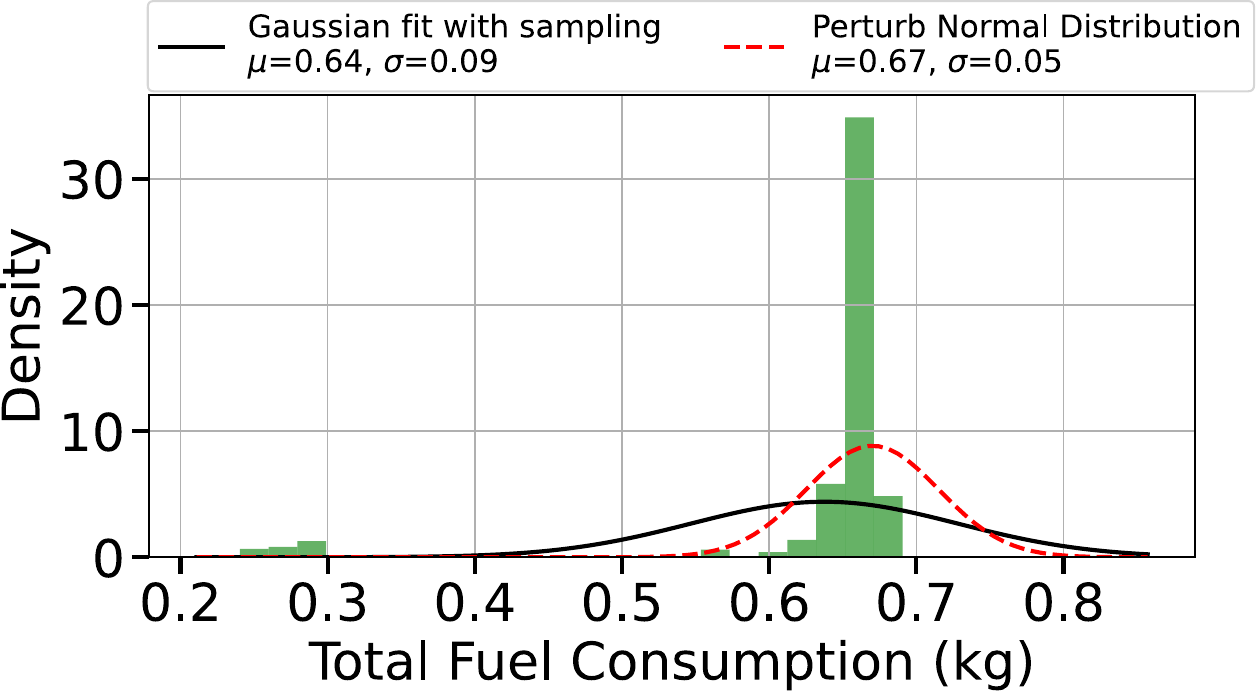}
 \caption{Perturbation Result for Total Fuel Consumption}
 \label{fig:perturb_active_fuel_SU}
 \end{subfigure}
 \caption{Perturbation results for CQI and Total Fuel Consumption of the Active-Control Net with Surrogate-based Environment}
 \label{fig:perturb_active_sidebyside_SU}
\end{figure*} 


 




 



When the surrogate models are used in place of the high-fidelity simulator for the Active-Control Net, 
Sobol's indices values are much less favorable than the high-fidelity simulator counterparts (see the Appendix). This may be attributed to the fact that the surrogate models do not have good enough accuracy to act as a reliable capture phase model. However, its performance for the perturbation method is promising as seen on Fig. \ref{fig:perturb_active_sidebyside_SU} (note that the axes are zoomed in for observability). In this test, the 1,000 samples were generated by the surrogate models as well, and the value of $\delta$ is set to be 0.001. Even though the errors are larger compared with the absolute value of the mean gained from the high-fidelity simulator in Figs. \ref{fig:perturb_active_cqi} and \ref{fig:perturb_active_fuel}, the results shown indicates that the surrogate models could be a good method to predict the gradients if their accuracies can be improved.

Furthermore, in the Appendix, Figs.~\ref{fig:sobol-scatter-all} and \ref{fig:sobol-scatter-all-fuel} show example Sobol indices for the entire input space (i.e., target states and \acp{MU} control), where the use of the surrogate model-based environment allows for a large number of sampling methods for this research. The time taken to execute 20,000 samples was only approx. 5 s; if the analysis were to be performed using the high-fidelity simulator, the data collection process would take at least 37,500 hrs. Fig. \ref{fig:sobol-scatter-all} shows that among all inputs, the target position, thrust angles $\psi$, and thrust magnitude $F_T$ are significantly more influential than others. As an indicator of capture success, this is reasonable because these variables determine whether the net is maneuvered properly to the target or not. Regarding the total fuel cost, Fig. \ref{fig:sobol-scatter-all-fuel} shows that $Z_D$, $\psi$, $\theta$, and $F_T$ are more influential in conjunction with the others, and that $\theta$ is especially important. This is also reasonable because $\theta$ determines how widely the net opens on the Z-Y plane in addition to how closely aligned each thrust vector is with the $-\mathbf{\hat{k}}$ direction. Therefore, the surrogate models -- although lacking precision in predicting the exact \ac{CQI} and total fuel cost -- can be used to formulate reasonable conclusions using the current \ac{UQ} framework, proving themselves to be promising and computationally efficient tools for the study of the tether-net system. The Appendix also presents the results of the same experiments discussed earlier in this section, but conducted on the surrogate environments, as shown in Fig.~\ref{fig:LF-passive}, ~\ref{fig:LF-passive-fuel} and ~\ref{fig:LF-passive-fuel-percept}.

\section{Conclusion}\label{sec:conclude}
In this paper, we propose a framework that conducts a sensitivity analysis and uncertainty propagation for the robotic tether-net system. Both the Fixed- and Active-Control Nets are evaluated with a high-fidelity simulation-based environment and a surrogate-model-based environment.
The results of this work have provided a deep understanding of the tether-net system from an uncertainty perspective, which highlights the most influential target position coordinate on the final \ac{CQI} and total fuel cost values. 
For the net using fixed thrust control, we found that variations of the coordinates $X_D$ and $Y_D$ have greater effects on the observed \ac{CQI} and total fuel consumption values in comparison to variations of $Z_D$. 
For the net with \ac{RL}-controlled thrust and perfect target position knowledge, $Z_D$ was found to be a lot more influential on the total fuel consumption value than the other two coordinates, while the \ac{CQI} value remained largely unaffected. 
Meanwhile, for the same system using imperfect target position knowledge, $X_D$ and $Y_D$ were found to have greater influences on the observed \ac{CQI}, while $Z_D$ had greater influence on the total fuel consumption. These observations can be uniquely helpful in informing the choice and performance features needed (e.g., what matters and what does not) for the sensing and estimation, while also allowing one to assess the reliability of the capture and fuel consumption calculations -- thereby allowing more rigorous yet compute-efficient robust design of the net control system.
The surrogate-model-based environment has also been shown to be promising in its ability to approximate the gradient of the high-fidelity simulation and has proven to be a computationally efficient method for conducting large-scale sampling to perform \ac{UQ}, which could not have been possible within a reasonable time frame using the high-fidelity simulator.
The results of this work also demonstrate that the perturbation method is capable of providing mean and variance estimates close to the corresponding values from Monte Carlo sampling.

\section*{Acknowledgments}
This work is supported under the CMMI Award numbered 2128578 from the National Science Foundation (NSF). The authors opinions, findings, and conclusions or recommendations expressed in this material do not necessarily reflect the views of the National Science Foundation.

\bibliographystyle{IEEEtran}
\bibliography{sample} 

\onecolumn
\section*{Appendix}
\begin{figure*}[htp!]
 \centering

 \begin{subfigure}[b]{0.9\linewidth}
 \centering
 \includegraphics[width=0.8\linewidth]{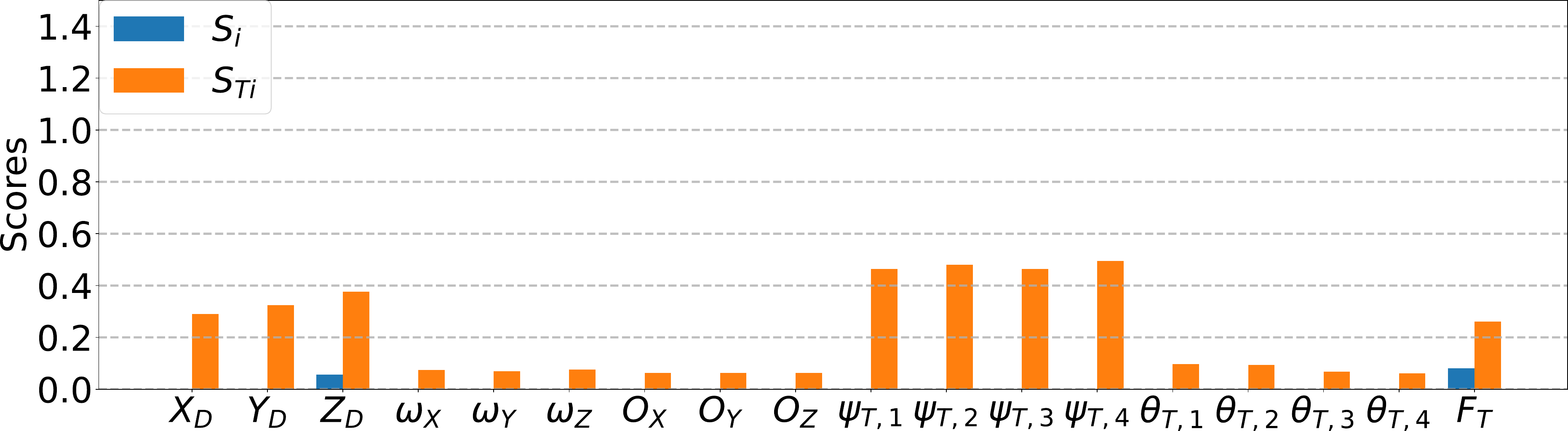}
 \caption{Sobol indices of CQI for all inputs with surrogate models}
 \label{fig:sobol-all}
 \end{subfigure}
 
 \vspace{0.1cm}

 \begin{subfigure}[b]{0.9\linewidth}
 \centering
 \includegraphics[width=1.0\linewidth]{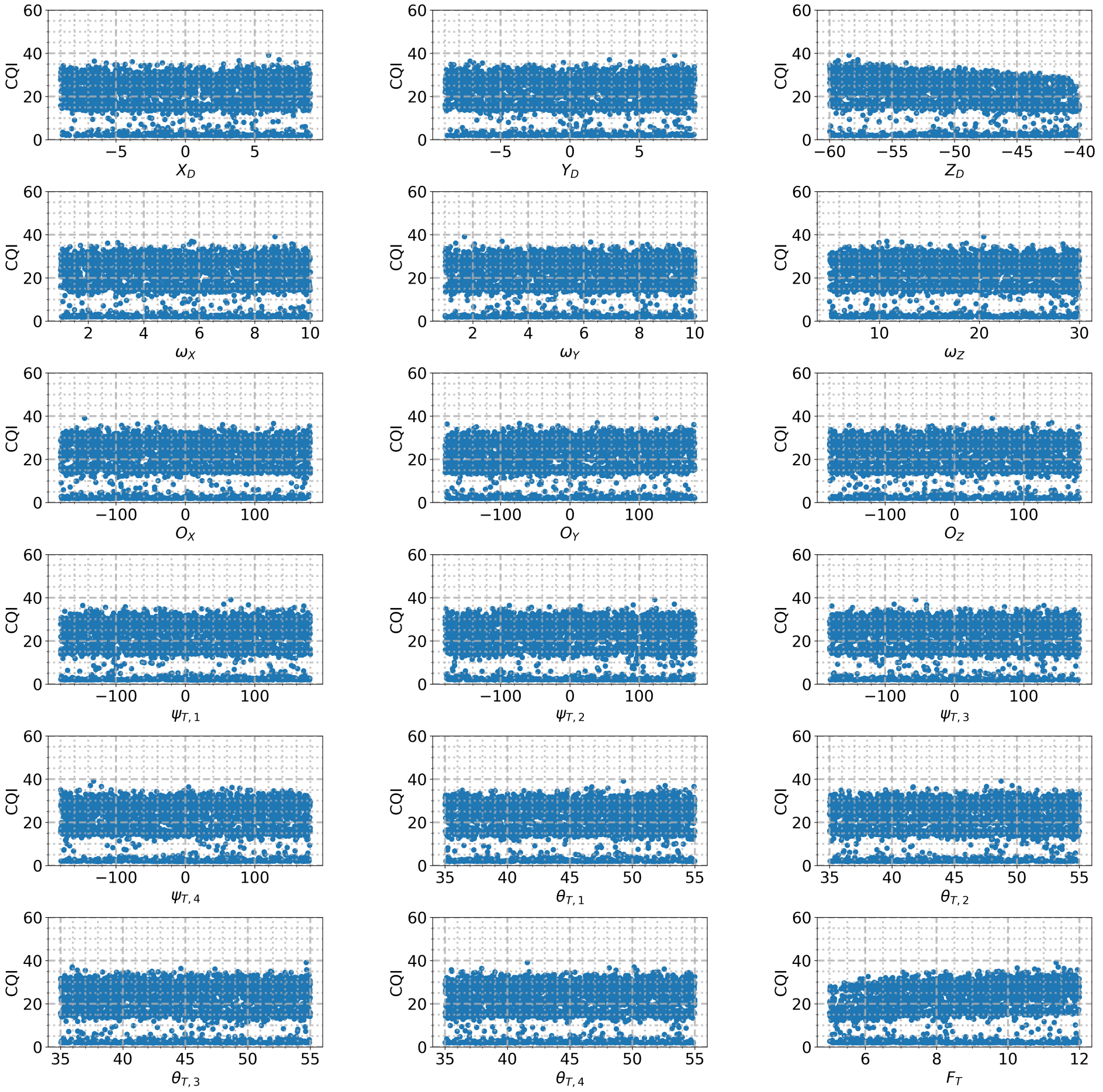}
 \caption{Scatter plot of CQI for all inputs with surrogate models}
 \label{fig:scatter-all}
 \end{subfigure}

 \caption{Sobol indices and scatter plots of CQI for all inputs using surrogate models}
 \label{fig:sobol-scatter-all}
\end{figure*}

\begin{figure*}[htp!]
 \centering

 \begin{subfigure}[b]{0.9\linewidth}
 \centering
 \includegraphics[width=0.8\linewidth]{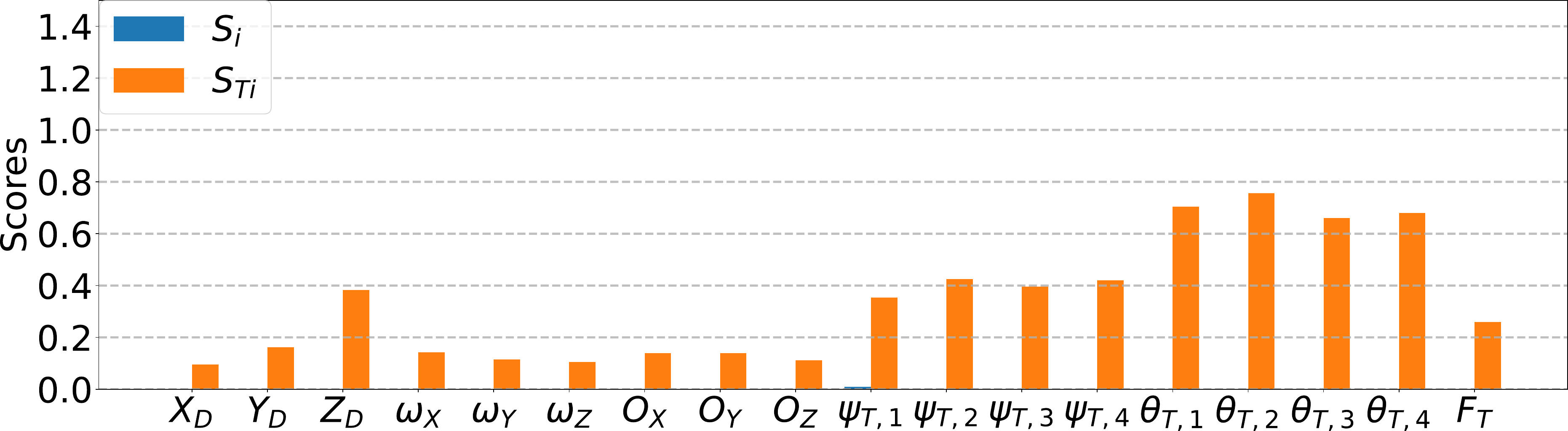}
 \caption{Sobol indices of Total Fuel Cost for all inputs with surrogate models}
 \label{fig:sobol-all-fuel}
 \end{subfigure}
 
 \vspace{0.1cm}

 \begin{subfigure}[b]{0.9\linewidth}
 \centering
 \includegraphics[width=1.0\linewidth]{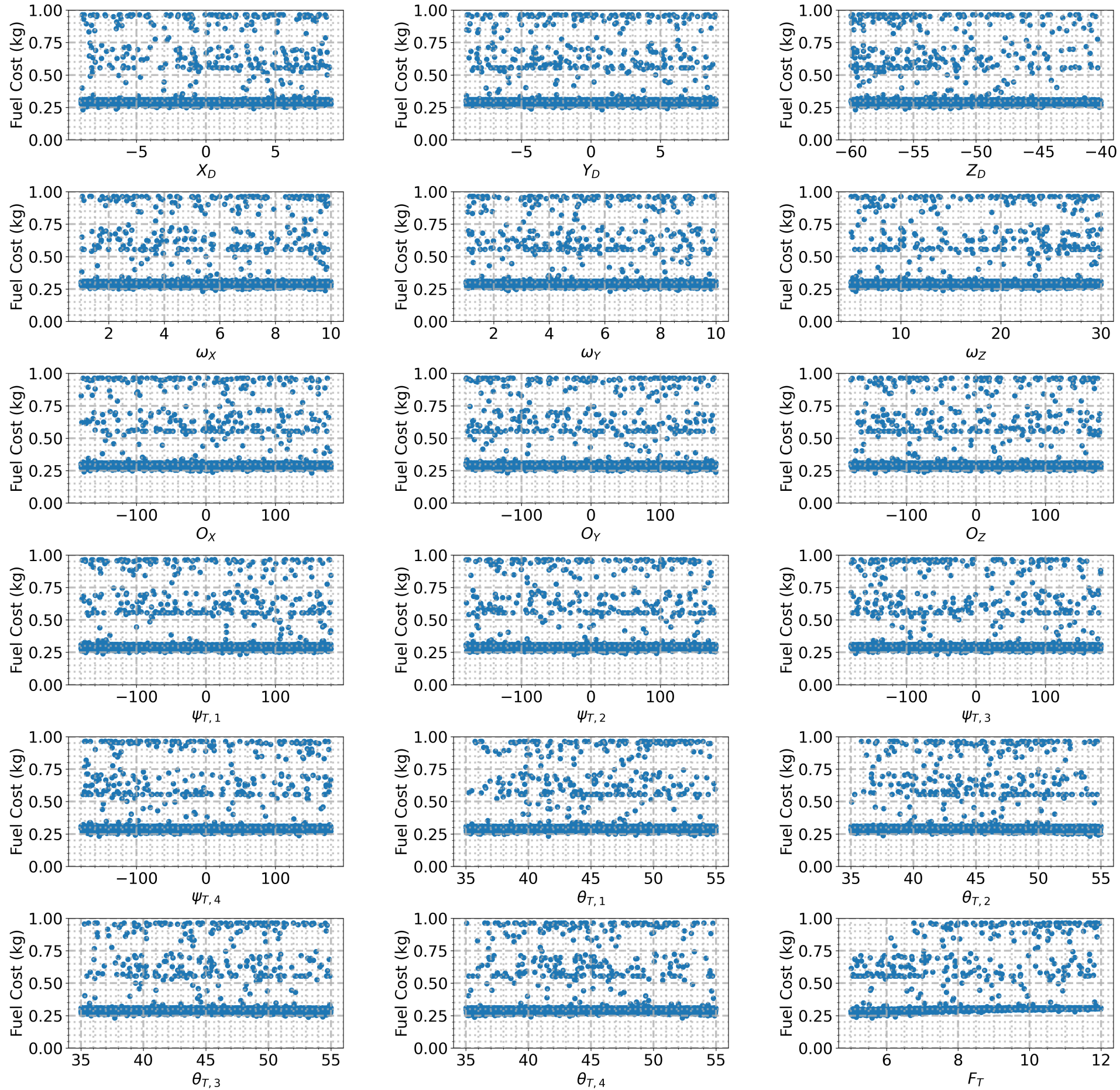}
 \caption{Scatter plot of Total Fuel Cost for all inputs with surrogate models}
 \label{fig:scatter-all-fuel}
 \end{subfigure}

 \caption{Sobol indices and scatter plots of Total Fuel Cost for all inputs using surrogate models}
 \label{fig:sobol-scatter-all-fuel}
\end{figure*}

\begin{figure*}[htp!]
 \centering
 \begin{subfigure}[t]{0.48\linewidth}
 \centering
 \includegraphics[width=\linewidth]{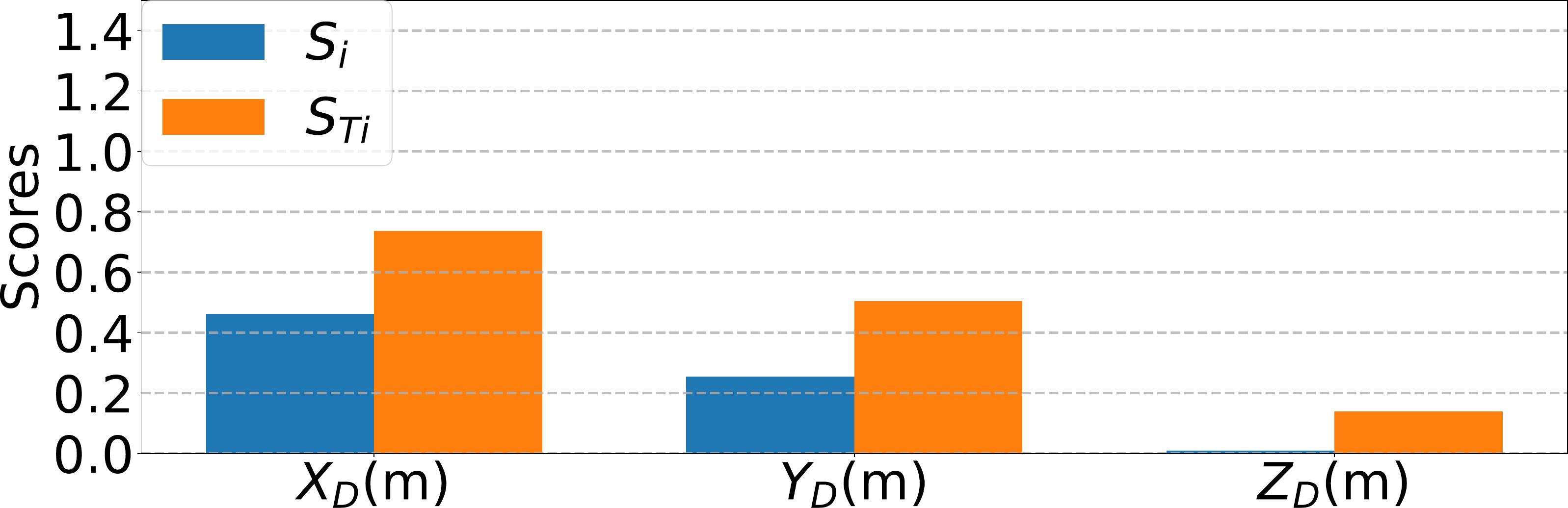}
 \caption{Sobol Indices of Target State (CQI)}
 \label{fig:LF sobol-Fixed-Control}
 \end{subfigure}
 \hfill
 \begin{subfigure}[t]{0.48\linewidth}
 \centering
 \includegraphics[width=\linewidth]{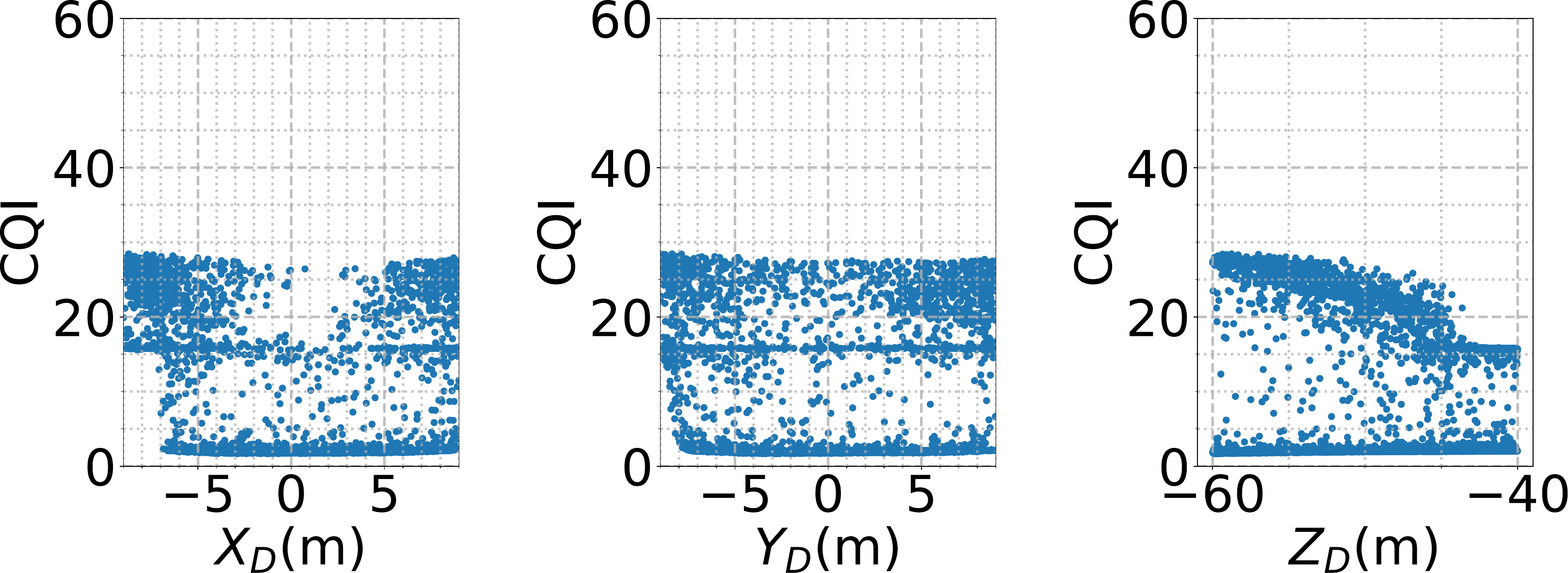}
 \caption{Scatterplot of Fixed-Control Net CQI}
 \label{fig:LF scatter-Fixed-Control}
 \end{subfigure}
 
 \vspace{1em} 

 \begin{subfigure}[t]{0.48\linewidth}
 \centering
 \includegraphics[width=\linewidth]{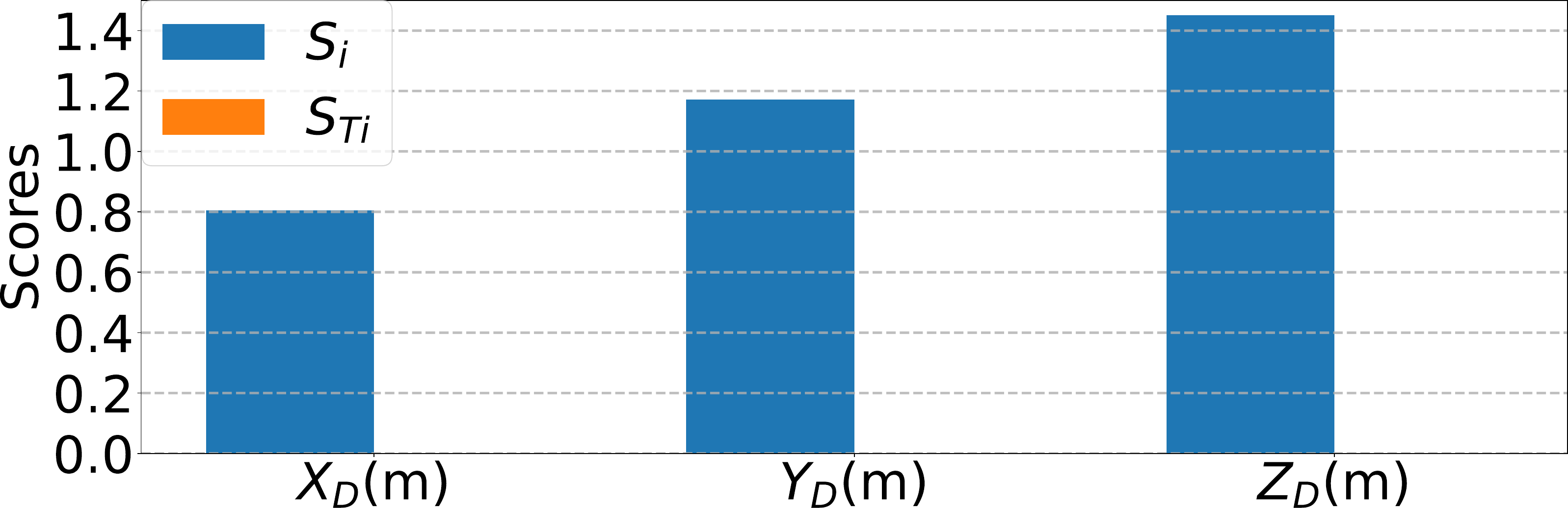}
 \caption{Sobol Indices of Target State (Total Fuel)}
 \label{fig:LF sobol-Fixed-Control-fuel}
 \end{subfigure}
 \hfill
 \begin{subfigure}[t]{0.48\linewidth}
 \centering
 \includegraphics[width=\linewidth]{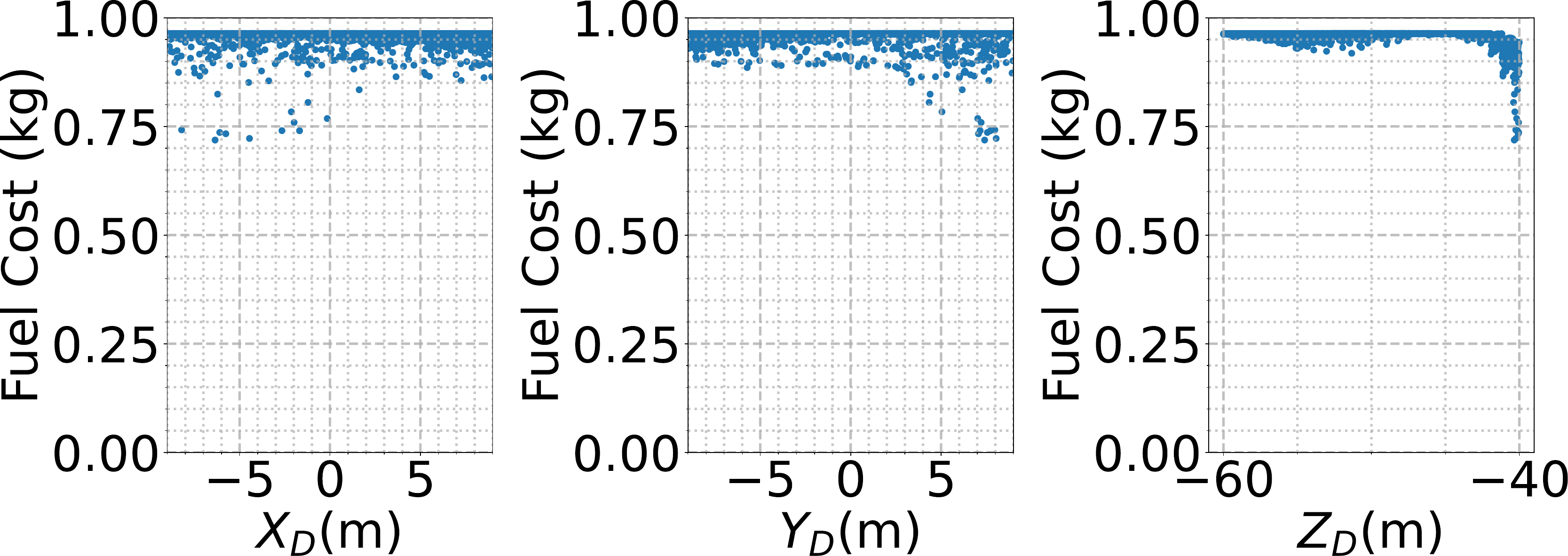}
 \caption{Scatterplot of Fixed-Control Net Total Fuel}
 \label{fig:LF scatter-Fixed-Control-fuel}
 \end{subfigure}

 \caption{Surrogate-based Environment - Comparison of Sobol Indices and Scatterplots for Fixed-Control Net on Two Metrics: CQI and Total Fuel}
 \label{fig:LF-passive}
\end{figure*}

\begin{figure*}[htp!]
 \centering
 \begin{subfigure}[t]{0.48\linewidth}
 \centering
 \includegraphics[width=\linewidth]{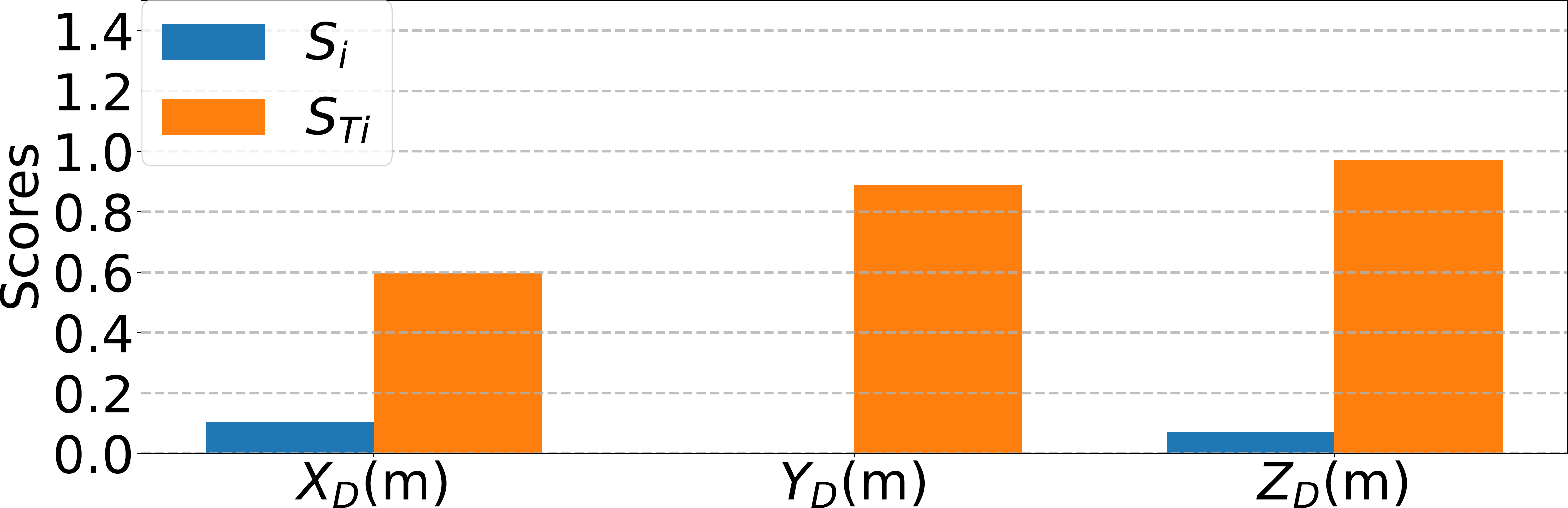}
 \caption{Sobol Indices of Target State (CQI)}
 \label{fig:LF sobol-Active-Control}
 \end{subfigure}
 \hfill
 \begin{subfigure}[t]{0.48\linewidth}
 \centering
 \includegraphics[width=\linewidth]{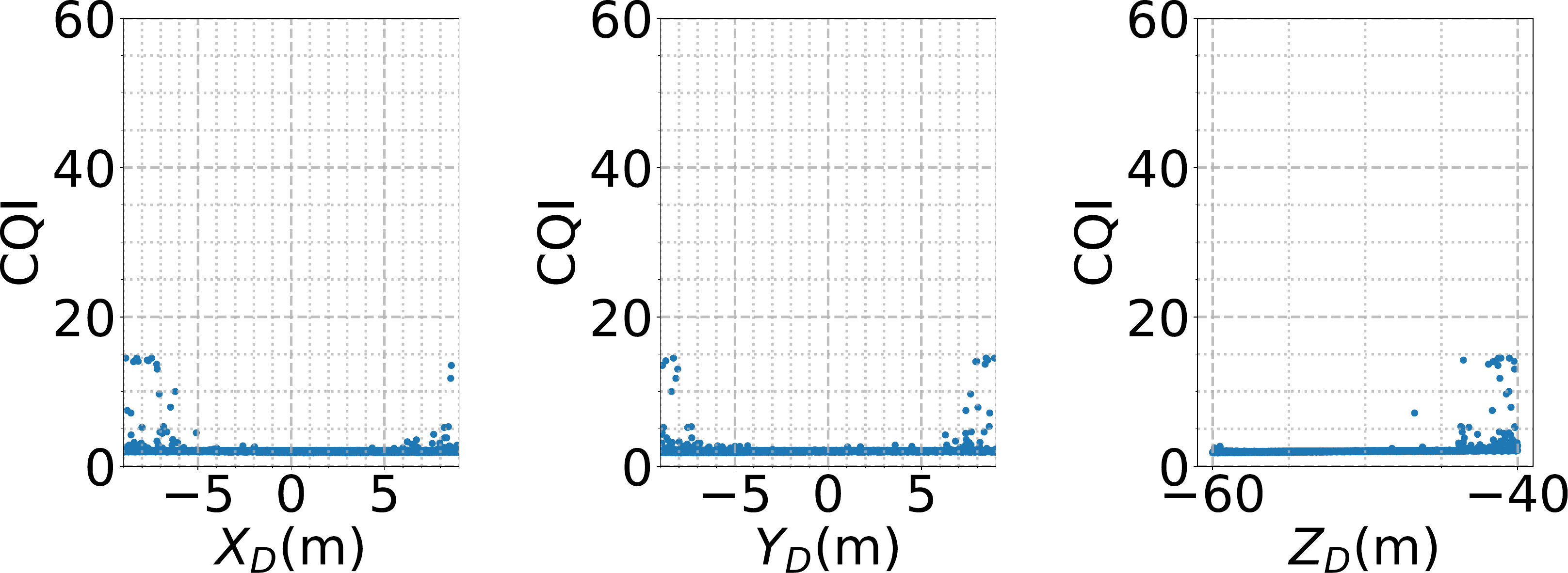}
 \caption{Scatterplot of Active-Control Net CQI}
 \label{fig:LF scatter-Active-Control}
 \end{subfigure}
 
 \vspace{1em} 

 \begin{subfigure}[t]{0.48\linewidth}
 \centering
 \includegraphics[width=\linewidth]{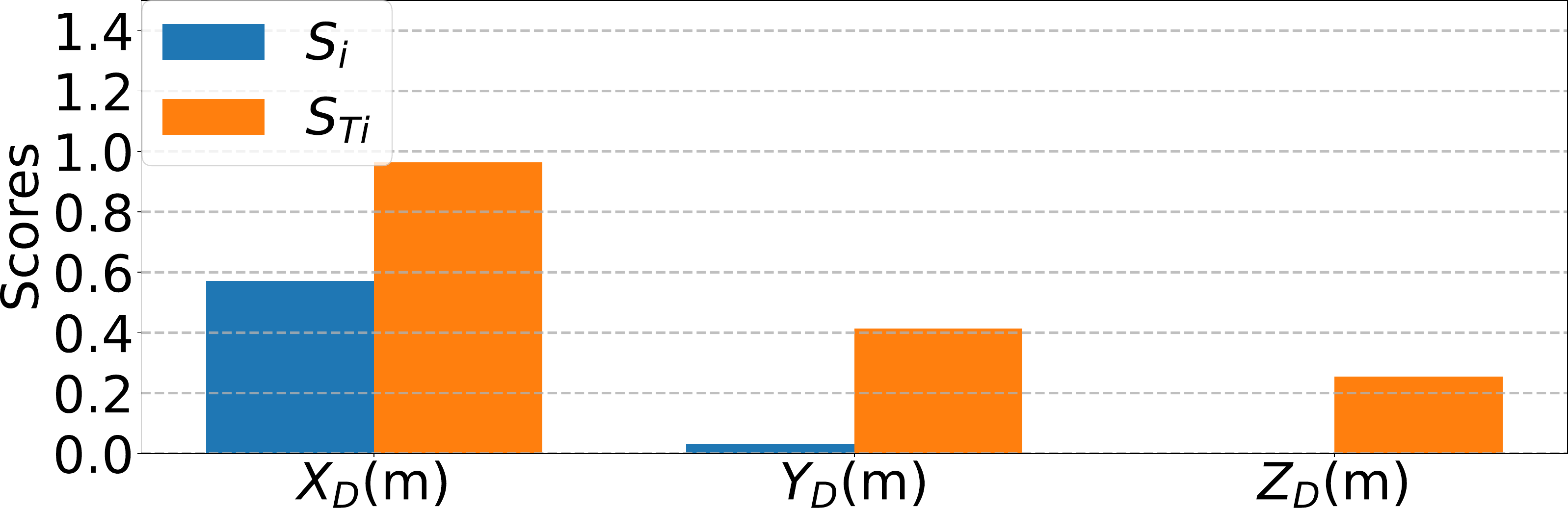}
 \caption{Sobol Indices of Target State (Total Fuel)}
 \label{fig:LF sobol-Active-Control-fuel}
 \end{subfigure}
 \hfill
 \begin{subfigure}[t]{0.48\linewidth}
 \centering
 \includegraphics[width=\linewidth]{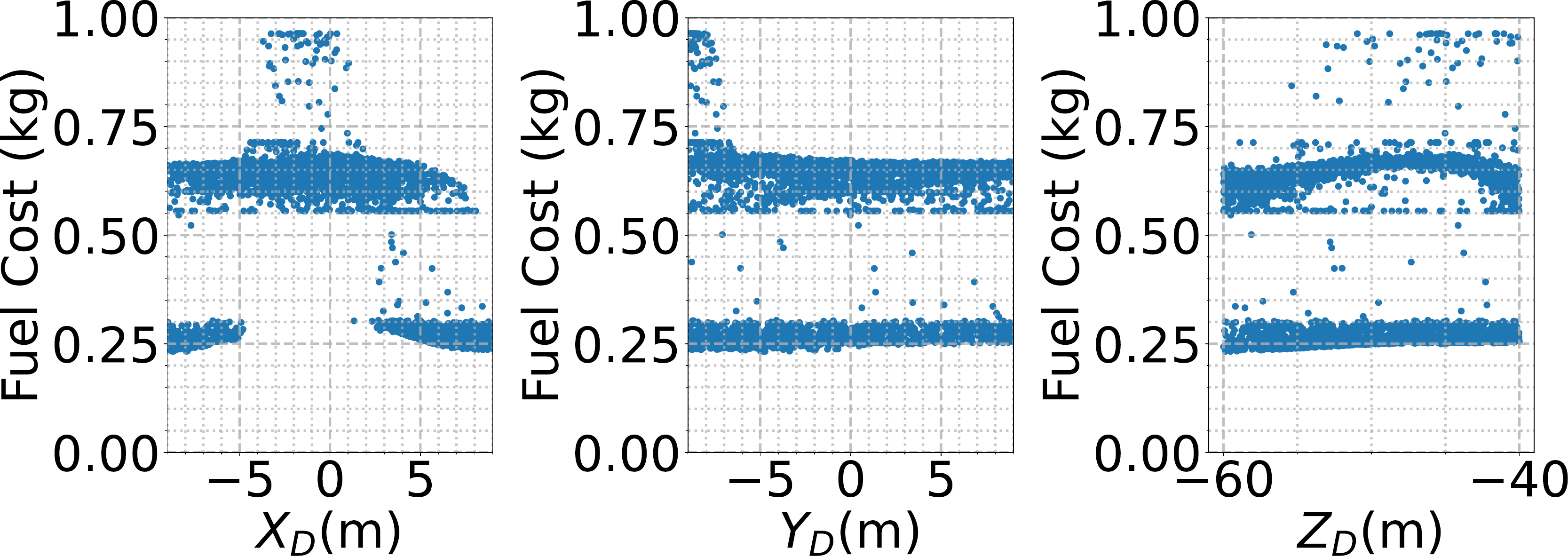}
 \caption{Scatterplot of Active-Control Net Total Fuel}
 \label{fig:LF scatter-Active-Control-fuel}
 \end{subfigure}

 \caption{Surrogate-based Environment - Comparison of Sobol Indices and Scatterplots for Active-Control Net on Two Metrics: CQI and Total Fuel}
 \label{fig:LF-passive-fuel}
\end{figure*}

\begin{figure*}[htp!]
 \centering
 \begin{subfigure}[t]{0.48\linewidth}
 \centering
 \includegraphics[width=\linewidth]{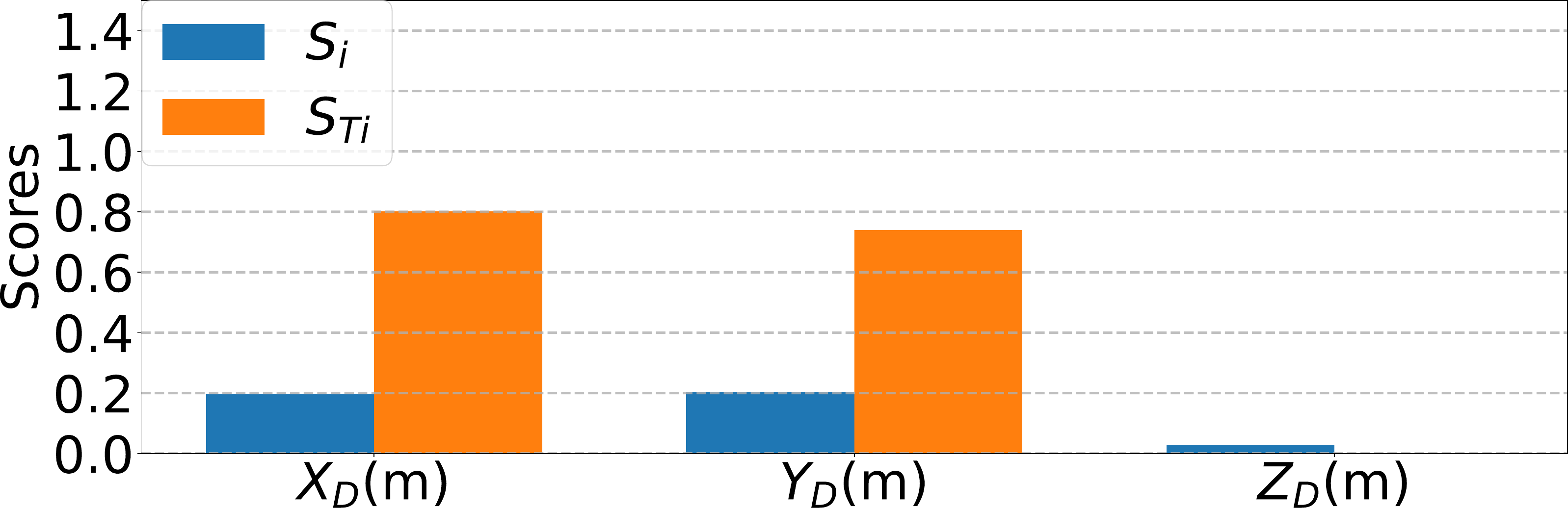}
 \caption{Sobol Indices of Target State (CQI)}
 \label{fig:LF sobol-Active-Control-percept}
 \end{subfigure}
 \hfill
 \begin{subfigure}[t]{0.48\linewidth}
 \centering
 \includegraphics[width=\linewidth]{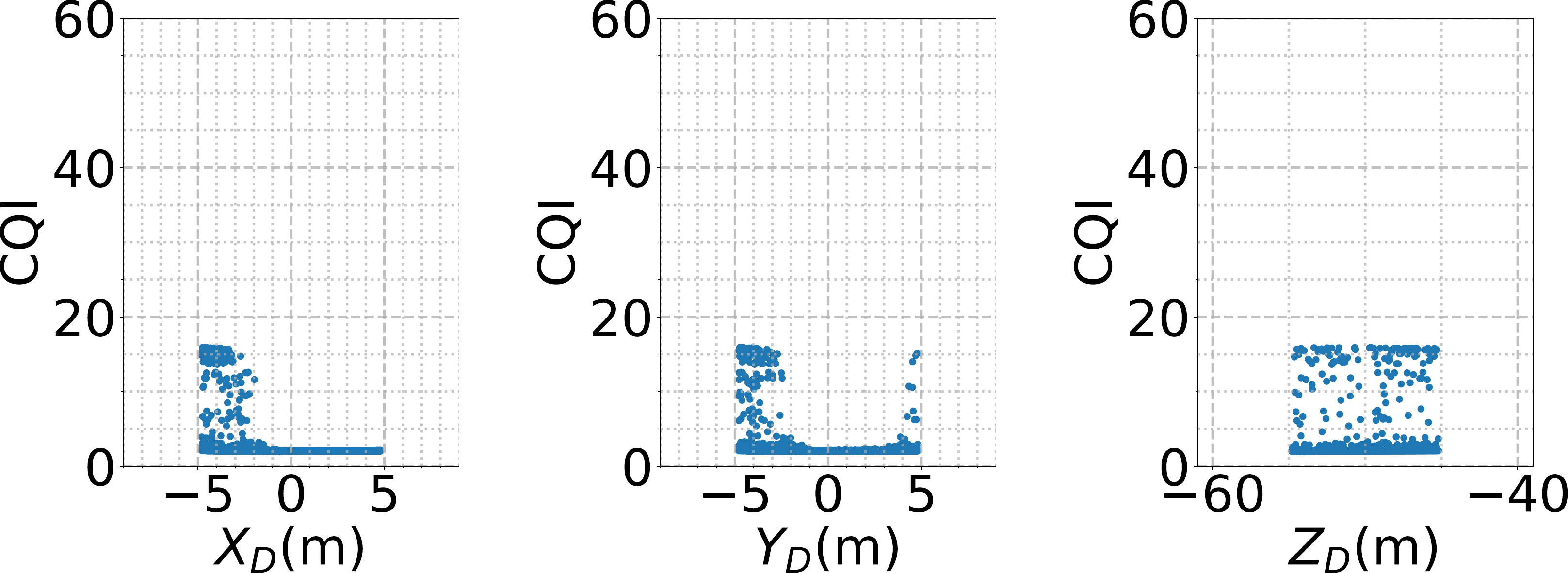}
 \caption{Scatterplot of Active-Control Net CQI}
 \label{fig:LF scatter-Active-Control-percept}
 \end{subfigure}
 
 \vspace{1em} 

 \begin{subfigure}[t]{0.48\linewidth}
 \centering
 \includegraphics[width=\linewidth]{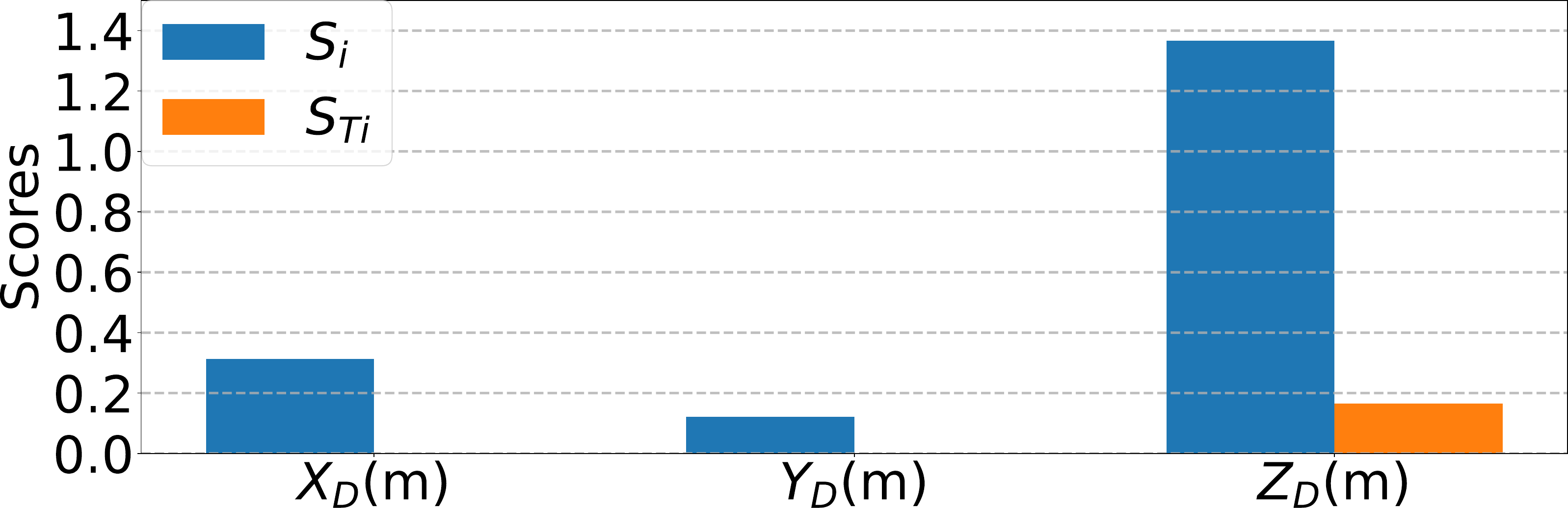}
 \caption{Sobol Indices of Target State (Total Fuel)}
 \label{fig:LF sobol-Active-Control-fuel-percept}
 \end{subfigure}
 \hfill
 \begin{subfigure}[t]{0.48\linewidth}
 \centering
 \includegraphics[width=\linewidth]{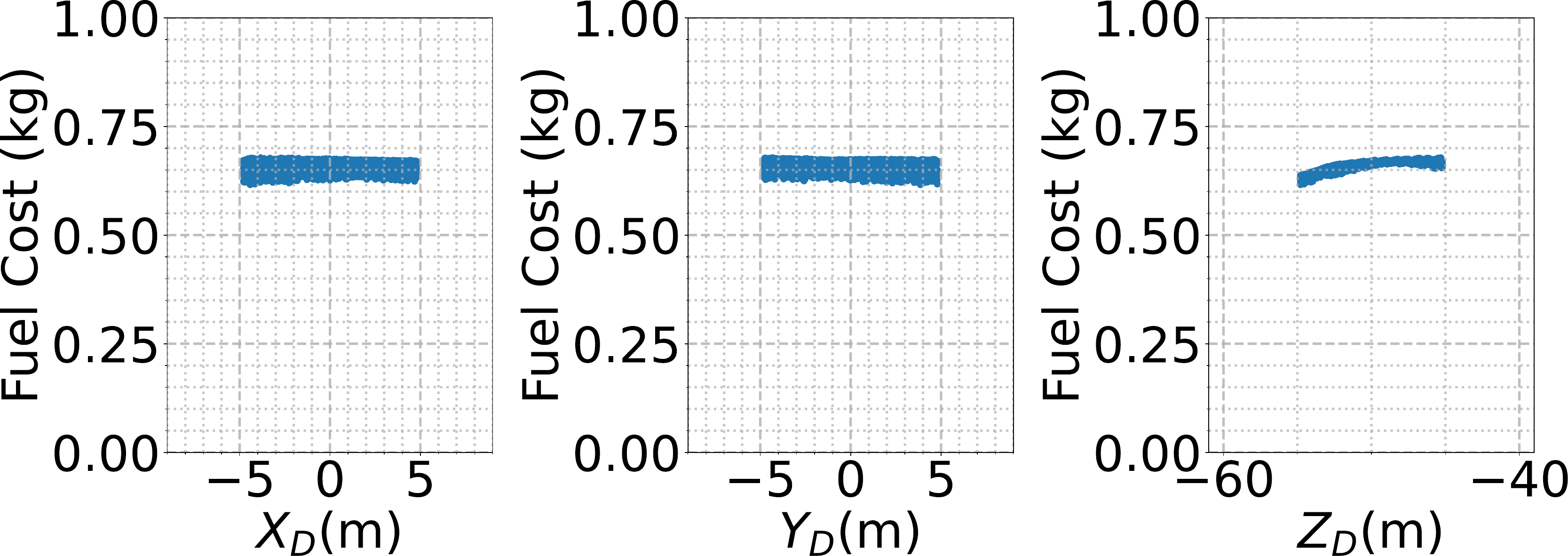}
 \caption{Scatterplot of Active-Control Net Total Fuel}
 \label{fig:LF scatter-Fixed-Control-fuel-percept}
 \end{subfigure}

 \caption{Surrogate-based Environment - Comparison of Sobol Indices and Scatterplots for Active-Control Net with Perception Noise on Two Metrics: CQI and Total Fuel}
 \label{fig:LF-passive-fuel-percept}
\end{figure*}

\end{document}